\documentstyle[epsf,11pt]{article}
\newcommand{\pr}{{Phys.\ Rev.\/}}

\newcommand{\np}{{Nucl.\ Phys.\/}}

\newcommand{\beq}{\begin{equation}}
\newcommand{\eeq}{\end{equation}}
\newcommand{\backn}{\begin{ack}}
\newcommand{\eackn}{\end{ack}}
\newcommand{\AmS}{{\protect\the\textfont2
  A\kern-.1667em\lower.5ex\hbox{M}\kern-.125emS}}

\title{Magnetic Monopoles and the Dual London Equation \\ in $SU(3)$ Lattice Gauge Theory\thanks{Supported in part by Fonds zur F\"oderung der wissenschaftlichen Forschung, P10745-PHY}} 

\author{Peter Skala, Manfried Faber, Martin Zach\\
Institut f\"ur Kernphysik, Technische Universit\"at Wien, A--1040 Vienna, Austria}

\begin{document}

\maketitle

\begin{abstract}
We propose a method for the determination of magnetic monopole currents in non-Abelian gauge theories which does not need a projection to Abelian degrees of freedom. With this definition we are able to determine the distribution of magnetic currents and electric fields for the gluonic flux tube between a pair of static charges. Further we check the validity of the Gauss law and the dual London equation in a gauge invariant formulation.
\end{abstract}

\section{Introduction}

The success of the hypothesis of the existence of quarks and the failure of the detection of quarks as isolated objects is a big challenge for Quantum Chromodynamics. The formulation of QCD on a lattice provides the possibility to show that quarks are confined by a linearly rising potential \cite{Creutz} the origin of which is the compression of the colourelectric flux between a pair of charges in a tube \cite{sommer}. The mechanism for the squeezing of the electric flux lines which originate in the colour charges is still a very interesting open question. One of the most promising conjectures is the hypothesis of 't Hooft \cite{Hooft} and Mandelstam \cite{mandelstam} that the QCD vacuum behaves dually to a superconductor which expels magnetic flux lines by the Meissner effect. 

The validity of this dual superconductor picture can be tested much more easily in compact QED which is an Abelian simplification of the non-Abelian theory of QCD by using $U(1)$ instead of $SU(3)$ parallel transporters. In compact QED one can define magnetic monopole currents \cite{degrand}. Magnetic charges form a dual Cooper pair condensate in the confined phase and are partly expelled \cite{Zach} by the dual Meissner effect from the region of the electric flux tube between a pair of electric charges. Curls of magnetic currents give rise to electric dipole moments. These electric dipoles align in the electric field of the charge pair. The electric field and the curl of the monopole currents are connected by the dual London equation \cite{haymaker}. The aligned magnetic currents form a thin solenoid around the electric flux tube.  

In order to test the dual superconductor picture of confinement in non-Abelian gauge theories one has to invent a prescription for the identification of magnetic monopole currents. It was 't Hooft's  idea \cite{thooft} that the magnetic monopoles may be identified in a $U(1)$ subgroup of the $SU(N)$ gauge group. This leads to the question of an appropriate procedure for the determination of the relevant degrees of freedom. Several partial gauge fixing procedures have been suggested which leave the largest Abelian subgroup unfixed. On the other hand a unique prescription is still missing. Moreover, in several articles problems appearing in the various gauge fixing procedures have been discussed \cite{green1}. 

In order to avoid the gauge fixing problem we suggest in this article an alternative prescription  \cite{Skala} for the determination of magnetic monopole currents in non-Abelian gauge theories. This description is an extension of the method suggested in \cite{Zach} for the identification of monopoles in $U(1)$ theory. We introduce our suggestion with a discussion of determination of the field strength and its connection to the Gauss law. Then we discuss the definition of gauge covariant monopole currents and finally we compare colourelectric fields and the curl of colourmagnetic monopole currents around a pair of colour charges and discuss the validity of the dual London equation.

\section{Field strength and the Gauss law}

In order to have a well settled definition of the field strength on the lattice we start our investigation with the discussion of the Gauss law for a pair of static colour charges. In classical Chromodynamics the Gauss law can be derived by a variation of the full QCD action with respect to the forth component $A_{4,a}$ of the vector potential, with $a$ being the colour index. To derive the Gauss law for Quantum Chromodynamics we use a version of Ehrenfest's theorem which is - for a gauge theory with gauge field $U$ - given by
\begin{equation}
\label{Ehren}
\langle {\cal O} \frac{\delta S}{\delta U}\rangle \; = \; \langle\frac{\delta {\cal O}}{\delta U}\rangle,
\end{equation}
with ${\cal O}$ being an arbitrary operator. In the path integral formalism (\ref{Ehren}) is a consequence of the translation invariance of the integration measure of the path integral. On a four-dimensional Euclidean hypercubic lattice $\Lambda$ of spacing $a$ with periodic boundary conditions in space and time direction the gauge fields are defined on links $(x,\mu)$ by parallel transporters
\begin{equation}
\label{para}
U_{\mu}(x) \; = \; e^{i\alpha^a_{\mu}(x) \, F_a} \qquad \qquad F_a = \frac{\lambda_a}{2}  , \qquad \alpha^a_{\mu}(x) = agA^a_\mu(x)  , \qquad x \in \Lambda  , \; \mu = 1,2,3,4
\end{equation}
where $\lambda_a$ denotes the Gell-Mann matrices and $g$ the QCD coupling constant. An action convenient for our purpose is the standard Wilson action
\begin{equation}
\label{action}
S[U] \; = \; \beta \sum_{x,\mu < \nu} \left( 1 - \frac{1}{3} \, \mbox{Re} \, \mbox{Tr} \, U_{\mu\nu}(x) \right), \;\;\;\; \beta = \frac{6}{g^2}
\end{equation}
with $U_{\mu\nu}(x)$ being the product of link variables around an elementary plaquette in $\mu\nu$-direction at lattice site $x$
\begin{equation}
\label{plaq}
U_{\mu\nu}(x) \; = \; U_{\mu}(x)U_{\nu}(x+\hat{\mu})U_{\mu}^\dagger(x+\hat{\nu})U_\nu^\dagger(x).
\end{equation}
Since we consider pure $SU(3)$ gauge theory with external static colour charges, the operator ${\cal O}$ in (\ref{Ehren}) is given by the Polyakov loop
\begin{equation}
\label{pol}
L(x_+) \, = \, \prod_{t=1}^{N_t} \, U_4(\vec{x}_+,t)
\end{equation}
describing the time evolution of a static charge at lattice site $x_+$. To derive the Gauss law at an arbitrary point $x$ the Polyakov loop (\ref{pol}) has to be parallel transported to $x$ by a Schwinger line $U(x,x_+)$ and reads
\begin{equation}
\label{pol2}
L^x(x_+) \, = \, U(x,x_+)L(x_+)U(x_+,x),
\end{equation}
with $U(x,x_+)$ being the product of link variables along a path connecting the lattice sites $x$ and $x_+$. For the variation in (\ref{Ehren}) we have to write
\begin{equation}
\label{var}
\frac{\delta}{\delta \alpha^a_4(x')} \; = \; \frac{\partial}{\partial \alpha^a_4(x)} \; \delta_{xx'}.
\end{equation}
Summation in colour space guarantees a gauge invariant formulation of the Gauss law and we receive for $x \neq x_-$ for (\ref{Ehren}) the following expression
\begin{eqnarray}
\label{rel1}
\hspace{-2cm}\lefteqn{\int {\cal D} [U] \, \frac{\partial S}{\partial\alpha^a_4(x)} \, \mbox{Tr}  \left( F_a  \, L^x(x_+) \right) \, \mbox{Tr} \, L^\ast(\vec{x}_-) \, e^{-S[U]} \, = } \nonumber \hspace{4cm} \\
&& =\,\int {\cal D} [U] \, \frac{\partial}{\partial\alpha^a_4(x)} \mbox{Tr}  \left( F_a  L^x(x_+) \, \right) \, \mbox{Tr} \, L^\ast(\vec{x}_-) \, e^{-S[U]}.    
\end{eqnarray}
The case $x=x_-$ is treated in section $4$. The approximation
\begin{equation}
\label{approx}
\frac{\partial U_4(x)}{\partial \alpha^a_4(x)} \approx i \, F_a \; U_4(x)
\end{equation}  
leads to
\begin{equation}
\label{rside}
i\, {\cal C}^2_{(f)}\; \int {\cal D} [U] \; \mbox{Tr}\, L(\vec{x}_+)\; \mbox{Tr}\, L^\ast(\vec{x}_-) \; e^{-S[U]} \; \delta_{x,x_+} 
\end{equation}
for the right-hand side of expression (\ref{rel1}), with
\begin{equation}
\label{casi}
{\cal C}^2_{(f)}\; := F_aF_a = \frac{4}{3},
\end{equation}
being the eigenvalue of the quadratic Casimir operator in the fundamental representation. To find a simplification for the left-hand side of (\ref{rel1}) we keep in mind that a  link variable $U_4(x)$ contributes to six space-time plaquettes $U_{i4}$. This we have to consider carrying out the derivative of the action $S$. Using
\begin{equation}
\label{proFa}
F_a \, \mbox{Tr} \left( F_aU_{\mu\nu}(x) \right) \; = \; \frac{1}{2} U_{\mu\nu}(x)_{(tl)}
\end{equation}
we obtain the following result for the Gauss law (\ref{rel1})
\begin{equation}
\label{gauss}
\frac{ \langle \, \mbox{Tr} \left\{ \left[ ga^2 \, \mbox{Div} \vec{E}(x) \right]^x \, L^x(x_+) \right\} \mbox{Tr} L^\ast(\vec{x}_-)\, \rangle }{ \langle \, \mbox{Tr}L(\vec{x}_+) \, \mbox{Tr} L^\ast(\vec{x}_-)\, \rangle } \; = \; i \frac{4}{3} \, g^2 \, \delta_{x,x_+}, \;\;\;\;\;\;\;\; x \neq x_-
\end{equation}
where we introduced the definition
\begin{equation}
\label{div}
\left(\mbox{Div} \vec{E}(x) \right)^x \, = \, \sum_{i=1}^{3}\, \left( E_i^x(x)-E_i^x(x-\hat{i}) \right)
\end{equation}
for the covariant divergence of the electric field strength. The electric field strength itself turns out to be the following hermitean traceless operator
\begin{equation}
\label{fields1}
ga^2 \, E_i^x(x) \, = \, \frac{1}{2i}\left(U_{4i}(x)-U_{4i}^\dagger(x)\right)_{(tl)}
\end{equation}
and is directly connected with the action (\ref{action}) used in our derivation of the Gauss law. If we want to know the field strength $E_i(x)$ at the lattice site $x+\hat{i}$, we have to carry out a parallel transport to the corresponding site which reads
\begin{equation}
\label{fields2}
E_i^{x+\hat{i}}(x) \, = \, U^\dagger_i(x) \, E_i^x(x) \, U_i(x)
\end{equation}
and is nontrivial for a non Abelian gauge theory. Of course, the field strength (\ref{fields1}) is an element of the $su(3)$-algebra and therefore as an $8$-vector in the algebra a gauge dependent quantity. But the correlation (\ref{gauss}) shows us how to calculate it as a gauge invariant quantity in lattice simulations. The term $\mbox{Tr} \left\{ \left[ga^2 \, \mbox{Div} \vec{E}(x_+)\right]^{x_+} \, L(x_+) \right\}$ in (\ref{gauss}) can be represented by the sum of worldlines
\begin{figure}[t]
\centerline{\begin{picture}(0,0)%
\special{psfile=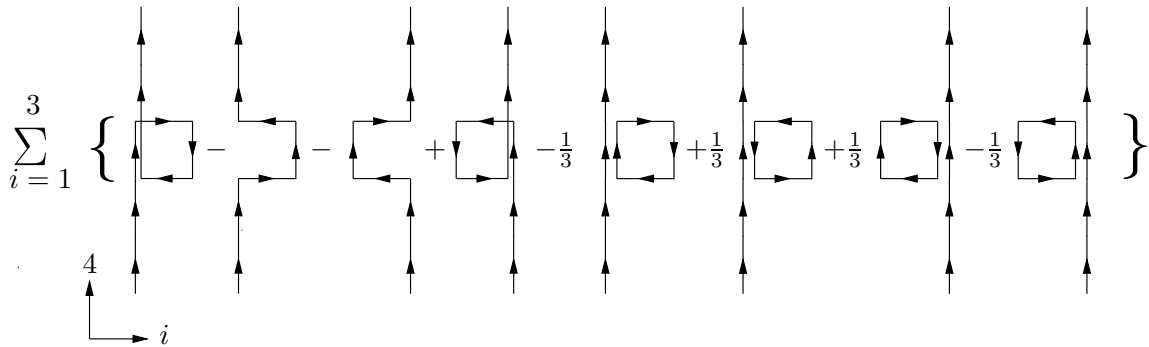}%
\end{picture}%
\setlength{\unitlength}{0.00020000in}%
\begingroup\makeatletter\ifx\SetFigFont\undefined
\def\x#1#2#3#4#5#6#7\relax{\def\x{#1#2#3#4#5#6}}%
\expandafter\x\fmtname xxxxxx\relax \def\y{splain}%
\ifx\x\y   
\gdef\SetFigFont#1#2#3{%
  \ifnum #1<17\tiny\else \ifnum #1<20\small\else
  \ifnum #1<24\normalsize\else \ifnum #1<29\large\else
  \ifnum #1<34\Large\else \ifnum #1<41\LARGE\else
     \huge\fi\fi\fi\fi\fi\fi
  \csname #3\endcsname}%
\else
\gdef\SetFigFont#1#2#3{\begingroup
  \count@#1\relax \ifnum 25<\count@\count@25\fi
  \def\x{\endgroup\@setsize\SetFigFont{#2pt}}%
  \expandafter\x
    \csname \romannumeral\the\count@ pt\expandafter\endcsname
    \csname @\romannumeral\the\count@ pt\endcsname
  \csname #3\endcsname}%
\fi
\fi\endgroup
\begin{picture}(29550,10716)(76,-10969)
\put(29300,-6136){\makebox(0,0)[lb]{\smash{\SetFigFont{36}{43.2}{rm}$\}$}}}
\put(5176,-5986){\makebox(0,0)[lb]{\smash{$-$}}}
\put(7951,-5986){\makebox(0,0)[lb]{\smash{$-$}}}
\put(10951,-5986){\makebox(0,0)[lb]{\smash{$+$}}}
\put( 76,-6200){\makebox(0,0)[lb]{\smash{\SetFigFont{20}{24}{rm}$\Sigma$}}}
\put( 0,-6800){\makebox(0,0)[lb]{\smash{$i=1$}}}
\put(450,-4850){\makebox(0,0)[lb]{\smash{$3$}}}
\put(3976,-10936){\makebox(0,0)[lb]{\smash{$i$}}}
\put(1951,-9061){\makebox(0,0)[lb]{\smash{$4$}}}
\put(2026,-6211){\makebox(0,0)[lb]{\smash{\SetFigFont{36}{43.2}{rm}$\{$}}}
\put(13876,-5986){\makebox(0,0)[lb]{\smash{$-\frac{1}{3}$}}}
\put(17851,-5986){\makebox(0,0)[lb]{\smash{$+\frac{1}{3}$}}}
\put(21500,-5986){\makebox(0,0)[lb]{\smash{$+\frac{1}{3}$}}}
\put(25200,-5986){\makebox(0,0)[lb]{\smash{$-\frac{1}{3}$}}}
\end{picture}}
\caption{\label{BildW}The worldlines of the 12 contributions in $\mbox{Tr} \left\{ \left[ga^2 \, \mbox{Div} \vec{E}(x_+)\right]^{x_+} \, L(x_+) \right\}$.}
\end{figure}
\renewcommand{\thefootnote}{\fnsymbol{footnote}}
shown in Fig.\ref{BildW}. The formulation of the Gauss law (\ref{gauss}) appears to be asymmetric with respect to the positions of quark and antiquark. According to our interpretation the Polyakov loop at $x_+$ defines a direction in the eight-dimensional group space. The average projection of $\mbox{Div}\vec{E}(x)$ to this direction is non zero and given by ${\cal C}^2_{(f)}$ for $x=x_+$ only. In other words the Polyakov loop is the source of an electric field which points in average in the direction given by the Polyakov loop itself. The antiquark in (\ref{gauss}) is only a ``spectator'' which is the necessary sink of the flux lines originating in $x_+$.\\
The right-hand side of the Gauss law (\ref{gauss}) corresponds to the square of the charge density of the considered colour sources. The charge density turns out to be imaginary as it should be in Euclidean space\footnote[2]{We use the following relation between Minkowski and Euclidean observables: All time components of 4-vectors transform like $x_4=ix_0$. In addition we use the convention $(\vec{E},\vec{B},\rho_m,\vec{J}_m)_{Euclidean}=(i\vec{E},\vec{B},\rho_m,-i\vec{J}_m)_{Minkowski}$.} \cite{charge} and is proportional to the eigenvalue of the quadratic Casimir operator ${\cal C}^2_{(f)}=\frac{4}{3}$ of a colour triplet which is a direct consequence of carrying out the trace in colour space in (\ref{rel1}).

The Gauss law can also be derived for a static colour octet. We consider a $3\otimes\bar{3}\,$-system at lattice site $x$ with generator
\begin{equation}
\label{g8}
F_a^{3\otimes\bar{3}} \; = \; F_a \otimes {\bf 1} + {\bf 1} \otimes \bar{F}_a
\end{equation}
and the Polyakov loops for this composed system
\begin{equation}
L^{3\otimes\bar{3}}(x) \; = \; L(x) \otimes L^\ast(x)
\end{equation}
and obtain in analogy to (\ref{rel1})
\begin{eqnarray}
\label{relO1}
\lefteqn{\int {\cal D} [U] \, \mbox{Tr}  \left[ F_a^{3\otimes\bar{3}} \frac{\partial S}{\partial\alpha^a_4(x)} \, L^{3\otimes\bar{3}}(x) \right] \, e^{-S[U]} \, = } \nonumber \hspace{4cm} \\
&& =\,\int {\cal D} [U] \, \mbox{Tr}  \left[  F_a^{3\otimes\bar{3}} \frac{\partial}{\partial\alpha^a_4(x)} \, L^{3\otimes\bar{3}}(x) \right] \, e^{-S[U]}.    
\end{eqnarray}
Since the uncharged singlet does not contribute, this leads to the Gauss law for the colour octet 
\begin{equation}
\label{gauss8}
\frac{\langle \; 2i \; \Im \left\{ \mbox{Tr} \left[(ga^2 \, \mbox{Div} \vec{E}(x))^xL(x)\right]\mbox{Tr} L^\ast(x)\right\}\; \rangle }{\langle \; \mbox{Tr}L(\vec{x}) \mbox{Tr} L^\ast(\vec{x}) - 1  \; \rangle } \; = \; i\; 3 g^2.
\end{equation}
As expected the colour charge on the right-hand side of (\ref{gauss8}) is proportional to the eigenvalue of the quadratic Casimir operator ${\cal C}^2_{(a)}=3$ in the adjoint representation.

\section{Definition of magnetic monopoles}

To identify magnetic monopoles in Abelian gauge theories DeGrand and Toussaint \cite{degrand} suggested to count the number of Dirac strings emanating from a three-dimensional cube. The Dirac strings carry a quantized magnetic flux of $\frac{2\pi}{e}$. The net flux for a closed surface defines the number of magnetic monopoles within a cube as a gauge invariant quantity. 't Hooft suggested to identify magnetic monopoles in non-Abelian gauge theories by a gauge fixing procedure where an appropriate operator in the adjoint representation of the gauge group is diagonalised. The monopoles are then defined as Dirac monopoles of the remaining $U(1)$-symmetry. The problem of this formalism is that the monopoles depend on the procedure used to define the Abelian projection.

\begin{figure}[t]
\centerline{\epsfxsize=6.5cm \epsfbox{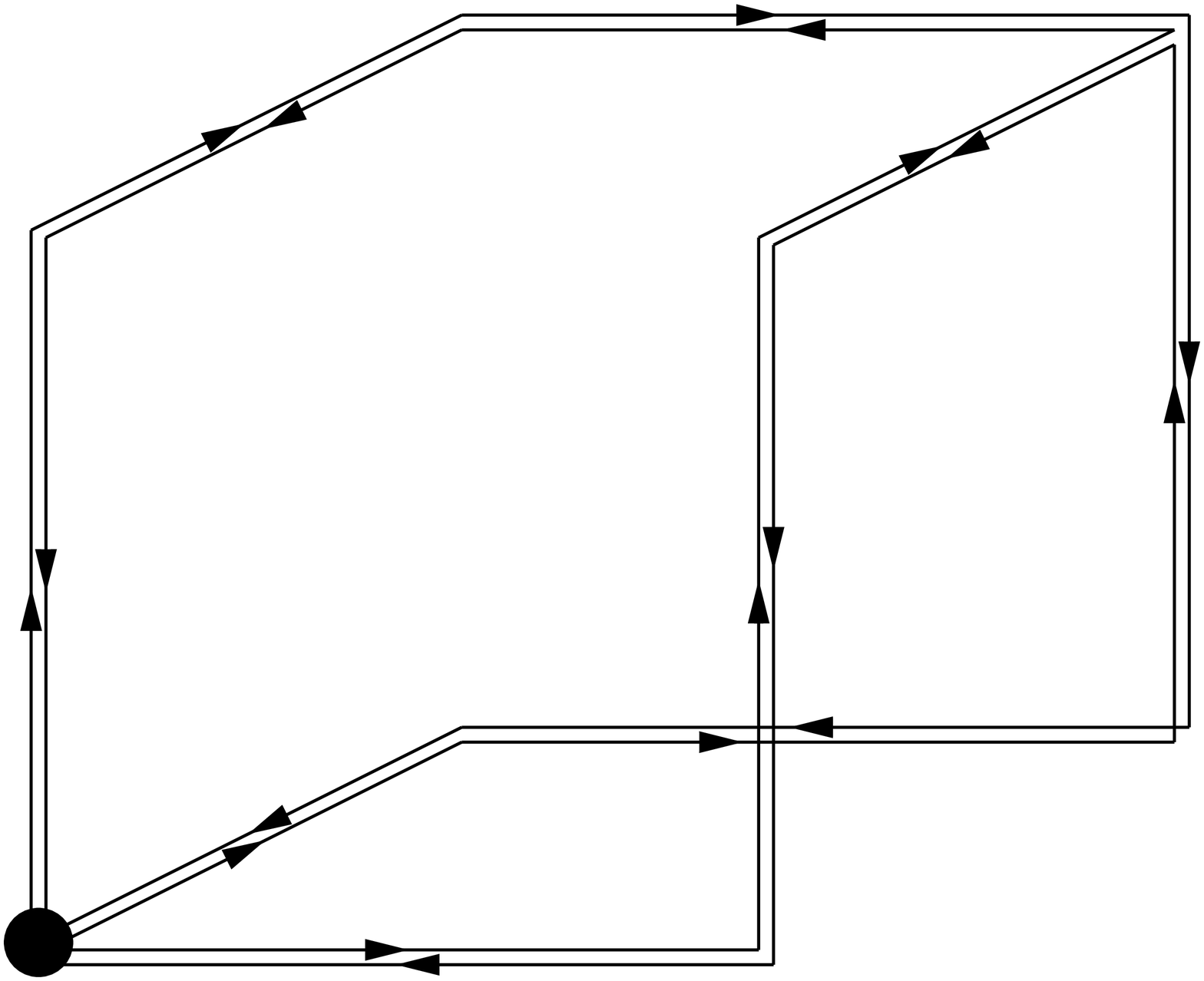}}
\caption{\label{Bianchi}Three pairs of plaquettes covering a monopole cube and measuring the flux out of the cube.}
\end{figure}

In \cite{Zach} an alternative definition of monopoles in Abelian gauge theories was suggested which uses the dual Maxwell equations. The monopole currents $J_{m,\mu}$ in this definition read on the lattice 
\begin{equation} \label{monop}
\varepsilon_{\mu \nu \rho \sigma} a^2 \sum_{\Box_i \in {\rm cube}(\nu \rho \sigma)} F_{\Box_i} = - J_{m,\mu} , \;\;\;\;\;\; \epsilon_{1234}=+1
\end{equation}
with $F_{\Box_i}$ being the field strength according to the definition
\begin{equation}
\label{U1field}
a^2eF_{\mu\nu} \; = \; \mbox{sin}\theta_{\mu\nu}
\end{equation}
where $a$ is the lattice spacing, $e$ the electric coupling and $\theta_{\mu\nu}$ the angle of an elementary plaquette in the $\mu\nu$-plane. In the following we describe a generalization of (\ref{monop}) to non-Abelian gauge theories as QCD. According to (\ref{fields1}) we define
\begin{equation}
\label{SU3field}
ga^2 \, F_{\mu\nu}^x(x) \; = \; \frac{1}{2i}\left( U_{\mu\nu}(x) - U_{\mu\nu}^\dagger(x) \right)_{(tl)}, \;\;\;\;\;\;\;\;\; \mu, \nu = 1,2,3,4
\end{equation}
i.e.
\begin{equation}
\label{emfield}
E_i^x(x) \, = \, F_{4i}^x(x), \;\;\;\;\;\;\;\;  B_i^x(x) \, = \, \frac{1}{2} \epsilon_{ijk} F_{jk}^x(x).   \;\;\;\;\;\;\;\; i,j,k = 1,2,3
\end{equation}
With the covariant derivative of the field strength
\begin{equation}
\label{covar}
D_{\nu}F_{\rho\sigma}^x(x) \, = \,F_{\rho\sigma}^x(x+\hat{\nu}) - F_{\rho\sigma}^x(x) \, = \, U_\nu(x) F_{\rho\sigma}^{x+\hat{\nu}}(x+\hat{\nu})U_\nu^\dagger(x) - F_{\rho\sigma}^x(x).
\end{equation}
the dual Maxwell equations in generalisation to (\ref{monop}) read
\begin{equation}
\label{dMax}
\frac{1}{2} \epsilon_{\mu\nu\rho\sigma} D_\nu F_{\rho\sigma}^x(x) \, = \, -J^x_{m,\mu}(x).
\end{equation}
\begin{figure}[t]
\centerline{\input{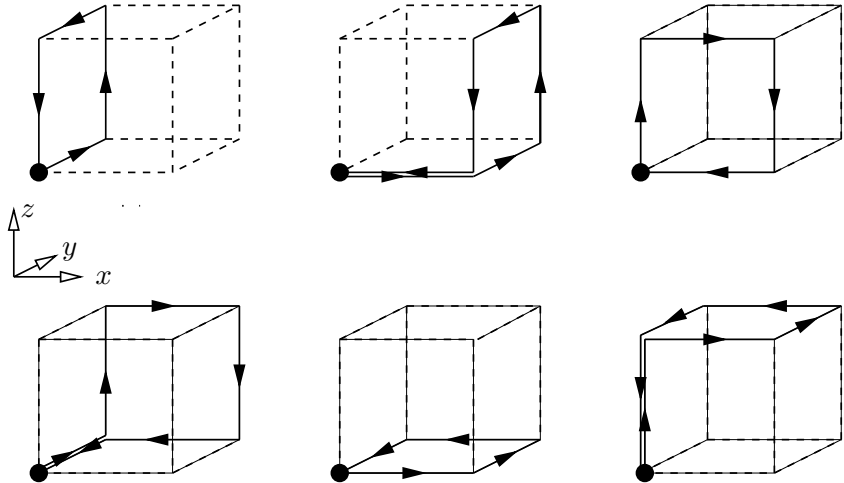}}
\caption{\label{moncur}The types of paths contributing to a three-dimensional spatial monopole cube, defining together with the reversed paths according to (23) the monopole current in time direction.}
\end{figure}
They define the colourmagnetic monopole current $J_{m,\mu}$. On the lattice each component of $J_{m,\mu}$ corresponds to a three-dimensional cube built of six plaquettes which measure the flux out of the cube, see Fig.\ref{moncur}. The covariant derivative (\ref{covar}) contains a parallel transport and therefore guarantees that all contributions to the total flux out of the cube are determined at the same lattice site $x$.\\
According to the Bianchi identities the magnetic currents (\ref{dMax}) should identically vanish. The lattice results reported below show that for finite lattice spacing this is obviously not the case. In the continuum the Bianchi identities are derived from the definition of the field strength via potentials. Also on the lattice they should be regained for ``infinitely small'' cubes. There may be two reasons why the magnetic currents do not vanish identically in lattice calculations: Bianchi identities  may be violated by Abelian monopoles and non-Abelian monopoles. Abelian monopoles are attached to Dirac strings, i.e. to $2\pi$-rotations of Abelian potentials. Such Abelian monopoles could appear in an arbitrary $U(1)$-subgroup of the full gauge group. They are pointlike and could be regarded as lattice artifacts. The non-Abelian monopoles could be associated with the non-Abelian nature of the gauge fields. To make this plausible we cover a cube by three pairs of plaquettes which are surrounded by paths as shown in Fig.\ref{Bianchi}. For the parallel transport along path ${\cal C}_i$ we write
\begin{equation}
\label{path}
{\cal U}_i = e^{i{\cal F}_i}, \;\;\;\;\; {\cal F }_i \in  su(3), \;\;\;\;\;\;\; i=1,2,3.
\end{equation}
It is obvious that
\begin{equation}
\label{unity}
e^{i{\cal F}_1}e^{i{\cal F}_2}e^{i{\cal F}_3} \; = \; \bf{1}.
\end{equation}
Nevertheless, according to the Baker-Campbell-Hausdorff formula we have in general
\begin{equation}
\label{BCH}
{\cal F}_1+{\cal F}_2+{\cal F}_3 \; \not=  \; 0,
\end{equation}
what leads to a non-vanishing contribution to the magnetic current which should vanish for infinitely small cubes. Therefore, such non-Abelian monopoles exist in the continuum limit only if they are extended objects. \\ 
\begin{figure}[hb]
\centerline{\epsfxsize=13cm \epsfbox{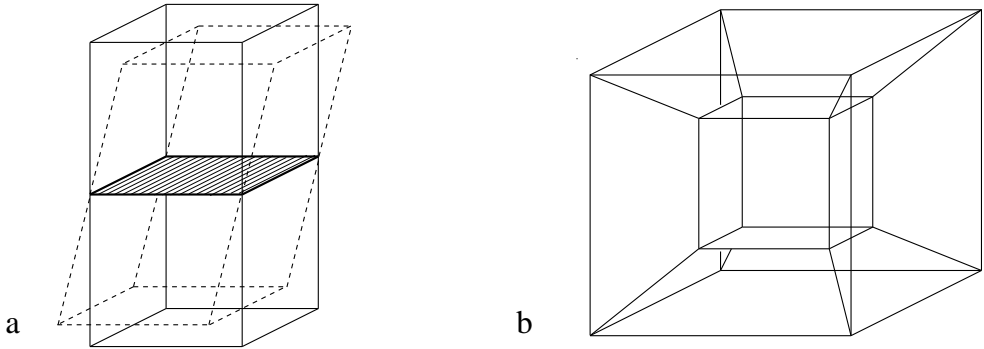}}
\caption{\label{Curcon}(a) One plaquette contributes to four monopole cubes or - seen from another point of view - the four monopole cubes contributing to a given component of the curl have one plaquette in common. (b) The eight monopole ``cubes'' contributing to the four-dimensional divergence of the monopole current on the lattice. Since every plaquette contributes twice, once with positive, once with negative sign, the current is conserved.}
\end{figure}
It is an interesting question if the current defined in (\ref{dMax}) is conserved. The flux through a given plaquette contributes to four adherent monopole cubes, see Fig.\ref{Curcon}a, which form the links of a dual plaquette on a dual lattice. The corresponding contribution to the magnetic current runs cyclically around this dual plaquette and is therefore conserved. We can discuss this problem also in another way. On the lattice there are eight magnetic current components contributing to the continuity equation. The corresponding eight cubes are depicted in Fig.\ref{Curcon}b, where the time evolution can be identified with a magnification of the small cube to the large cube. It can be easily seen that every plaquette contributes to two cubes with positive and negative sign respectively. If one guarantees that both contributions of a given plaquette are transported on the same path to the given reference point the sum of the $48$ contributions is zero. 

Since our aim is to verify the validity of the dual London equation, we need to define not only the colourmagnetic monopole currents  $J_{m,\nu}$ (\ref{dMax}) but also the curl of the monopole current. We use the following  discretisation of the curl
\begin{equation}
\label{rot}
\left( \mbox{curl} \vec{J}_m \right)_i^x (x) \, = \, \epsilon_{ijk} \left( D_j J_{m,k} \right)^x \, = \, \epsilon_{ijk} \left( J^x_{m,k}(x)-J^x_{m,k}(x-\hat{j}) \right) \;\;\;\;\;\;\;\;\; i,j,k = 1,2,3
\end{equation}
with $D_j$ being the covariant derivative as defined in (\ref{covar}), but we simultaneously take into account that the field strength contributions to the current on the right-hand side of (\ref{rot}) have to be transported along the shortest path to the reference site $x$ of the curl. For the further discussion of the London equation we want to emphasize that on the lattice the four monopole cubes contributing according to (\ref{rot}) to a given component of the curl have one plaquette in common, as shown in Fig.\ref{Curcon}a. This leads to a correspondence of the operators $( \mbox{curl} \vec{J}_m )_i^x (x) \longleftrightarrow E_i^x(x)$ which refer both to the same lattice site $x$.

We have now defined all necessary ingredients for a numerical determination of the electric field distribution and the distribution of the curl of the magnetic current and to test in the following section the validity of the dual London equation for the gluonic vacuum with external static colour charges.

\section{Correlation functions and numerical results}

\begin{figure}[h]
\centerline{\input{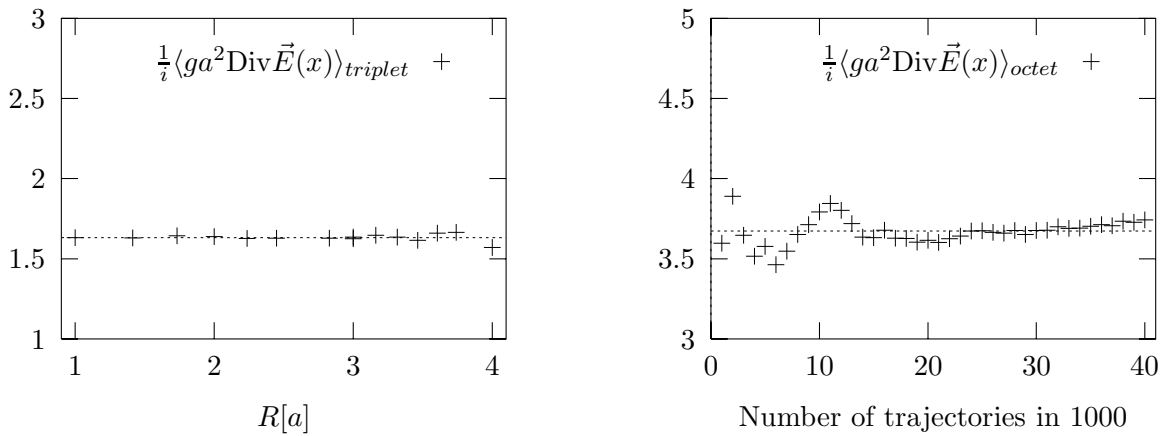}
\setlength{\unitlength}{0.1bp}
\special{!
/gnudict 40 dict def
gnudict begin
/Color false def
/Solid false def
/gnulinewidth 2.000 def
/vshift -33 def
/dl {10 mul} def
/hpt 31.5 def
/vpt 31.5 def
/M {moveto} bind def
/L {lineto} bind def
/R {rmoveto} bind def
/V {rlineto} bind def
/vpt2 vpt 2 mul def
/hpt2 hpt 2 mul def
/Lshow { currentpoint stroke M
  0 vshift R show } def
/Rshow { currentpoint stroke M
  dup stringwidth pop neg vshift R show } def
/Cshow { currentpoint stroke M
  dup stringwidth pop -2 div vshift R show } def
/DL { Color {setrgbcolor Solid {pop []} if 0 setdash }
 {pop pop pop Solid {pop []} if 0 setdash} ifelse } def
/BL { stroke gnulinewidth 2 mul setlinewidth } def
/AL { stroke gnulinewidth 2 div setlinewidth } def
/PL { stroke gnulinewidth setlinewidth } def
/LTb { BL [] 0 0 0 DL } def
/LTa { AL [1 dl 2 dl] 0 setdash 0 0 0 setrgbcolor } def
/LT0 { PL [] 0 1 0 DL } def
/LT1 { PL [4 dl 2 dl] 0 0 1 DL } def
/LT2 { PL [2 dl 3 dl] 1 0 0 DL } def
/LT3 { PL [1 dl 1.5 dl] 1 0 1 DL } def
/LT4 { PL [5 dl 2 dl 1 dl 2 dl] 0 1 1 DL } def
/LT5 { PL [4 dl 3 dl 1 dl 3 dl] 1 1 0 DL } def
/LT6 { PL [2 dl 2 dl 2 dl 4 dl] 0 0 0 DL } def
/LT7 { PL [2 dl 2 dl 2 dl 2 dl 2 dl 4 dl] 1 0.3 0 DL } def
/LT8 { PL [2 dl 2 dl 2 dl 2 dl 2 dl 2 dl 2 dl 4 dl] 0.5 0.5 0.5 DL } def
/P { stroke [] 0 setdash
  currentlinewidth 2 div sub M
  0 currentlinewidth V stroke } def
/D { stroke [] 0 setdash 2 copy vpt add M
  hpt neg vpt neg V hpt vpt neg V
  hpt vpt V hpt neg vpt V closepath stroke
  P } def
/A { stroke [] 0 setdash vpt sub M 0 vpt2 V
  currentpoint stroke M
  hpt neg vpt neg R hpt2 0 V stroke
  } def
/B { stroke [] 0 setdash 2 copy exch hpt sub exch vpt add M
  0 vpt2 neg V hpt2 0 V 0 vpt2 V
  hpt2 neg 0 V closepath stroke
  P } def
/C { stroke [] 0 setdash exch hpt sub exch vpt add M
  hpt2 vpt2 neg V currentpoint stroke M
  hpt2 neg 0 R hpt2 vpt2 V stroke } def
/T { stroke [] 0 setdash 2 copy vpt 1.12 mul add M
  hpt neg vpt -1.62 mul V
  hpt 2 mul 0 V
  hpt neg vpt 1.62 mul V closepath stroke
  P  } def
/S { 2 copy A C} def
end
}
\begin{picture}(2339,1511)(0,0)
\special{"
gnudict begin
gsave
50 50 translate
0.100 0.100 scale
0 setgray
/Helvetica findfont 100 scalefont setfont
newpath
-500.000000 -500.000000 translate
LTa
480 251 M
0 1209 V
LTb
480 251 M
63 0 V
1613 0 R
-63 0 V
480 553 M
63 0 V
1613 0 R
-63 0 V
480 856 M
63 0 V
1613 0 R
-63 0 V
480 1158 M
63 0 V
1613 0 R
-63 0 V
480 1460 M
63 0 V
1613 0 R
-63 0 V
480 251 M
0 63 V
0 1146 R
0 -63 V
889 251 M
0 63 V
0 1146 R
0 -63 V
1298 251 M
0 63 V
0 1146 R
0 -63 V
1706 251 M
0 63 V
0 1146 R
0 -63 V
2115 251 M
0 63 V
0 1146 R
0 -63 V
480 251 M
1676 0 V
0 1209 V
-1676 0 V
480 251 L
LT0
1913 1297 A
521 612 A
562 789 A
603 642 A
644 563 A
684 600 A
725 531 A
766 582 A
807 645 A
848 682 A
889 730 A
930 762 A
971 736 A
1011 686 A
1052 635 A
1093 633 A
1134 661 A
1175 631 A
1216 630 A
1257 616 A
1298 623 A
1338 615 A
1379 629 A
1420 639 A
1461 658 A
1502 659 A
1543 653 A
1584 650 A
1625 660 A
1665 645 A
1706 660 A
1747 661 A
1788 674 A
1829 668 A
1870 669 A
1911 676 A
1952 682 A
1992 678 A
2033 695 A
2074 691 A
2115 700 A
LT3
480 658 M
17 0 V
17 0 V
17 0 V
17 0 V
17 0 V
17 0 V
17 0 V
16 0 V
17 0 V
17 0 V
17 0 V
17 0 V
17 0 V
17 0 V
17 0 V
17 0 V
17 0 V
17 0 V
17 0 V
17 0 V
17 0 V
16 0 V
17 0 V
17 0 V
17 0 V
17 0 V
17 0 V
17 0 V
17 0 V
17 0 V
17 0 V
17 0 V
17 0 V
17 0 V
17 0 V
16 0 V
17 0 V
17 0 V
17 0 V
17 0 V
17 0 V
17 0 V
17 0 V
17 0 V
17 0 V
17 0 V
17 0 V
17 0 V
17 0 V
16 0 V
17 0 V
17 0 V
17 0 V
17 0 V
17 0 V
17 0 V
17 0 V
17 0 V
17 0 V
17 0 V
17 0 V
17 0 V
17 0 V
16 0 V
17 0 V
17 0 V
17 0 V
17 0 V
17 0 V
17 0 V
17 0 V
17 0 V
17 0 V
17 0 V
17 0 V
17 0 V
17 0 V
16 0 V
17 0 V
17 0 V
17 0 V
17 0 V
17 0 V
17 0 V
17 0 V
17 0 V
17 0 V
17 0 V
17 0 V
17 0 V
17 0 V
16 0 V
17 0 V
17 0 V
17 0 V
17 0 V
17 0 V
17 0 V
17 0 V
stroke
grestore
end
showpage
}
\put(1793,1297){\makebox(0,0)[r]{$\frac{1}{i} \langle ga^2\mbox {Div}\vec{E}(x)\rangle_{octet}$}}
\put(1318,-49){\makebox(0,0){Number of trajectories in $1000$}}
\put(2115,151){\makebox(0,0){40}}
\put(1706,151){\makebox(0,0){30}}
\put(1298,151){\makebox(0,0){20}}
\put(889,151){\makebox(0,0){10}}
\put(480,151){\makebox(0,0){0}}
\put(420,1460){\makebox(0,0)[r]{5}}
\put(420,1158){\makebox(0,0)[r]{4.5}}
\put(420,856){\makebox(0,0)[r]{4}}
\put(420,553){\makebox(0,0)[r]{3.5}}
\put(420,251){\makebox(0,0)[r]{3}}
\end{picture}}
\vspace{0.6cm}
\caption{\label{gauss3}The Gauss law for a colour triplet as a function of the distance $R[a]$ between the charges (left figure, see Eq. (13)) and for a colour octet as a function of the number of evaluated gauge field configurations (right figure, see Eq. (20)). In both cases the theoretical prediction is plotted as a dotted line and agrees very well with the numerical values.} 
\end{figure}

In this section we would like to present our numerical results. All simulations were done on a Euclidean $8^3 \times 2$-lattice in the confinement phase at $\beta=4.9$. The gauge field configurations were generated with a Hybrid-Monte-Carlo-algorithm.\\
First let us begin with the ``measurements'' concerning the verification of the Gauss law for a colour triplet respectively a colour octet. The left diagram of Fig.\ref{gauss3} shows the results of evaluating the correlation function (\ref{gauss})
of $40.000$ gauge field configurations for different distances $R[a]$ of the colour charges. The theoretical prediction is plotted as a dotted line and agrees very well with the numerical results of the left-hand side of (\ref{gauss}). The right diagram of Fig.\ref{gauss3} shows the numerical value of the correlation function (\ref{gauss8}) as a function of the number of evaluated gauge field configurations. The results agree with the theoretical value.

\begin{figure}
\vspace{-0.2cm}
\centerline{\input{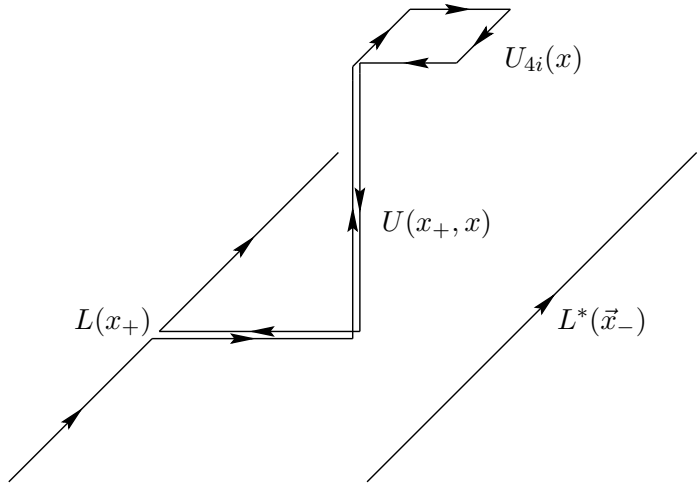}}
\vspace{0.3cm}
\caption{\label{BildE}A Schwinger line $U(x_+,x)$ connects the field strength $E_i(x)=\frac{1}{2i}\left( U_{4i}(x)-U_{4i}^\dagger(x) \right)_{(tl)}$ and the Polyakov loop $L(x+)$ to guarantee gauge invariance.} 
\end{figure}

Now let's turn to the correlation functions measuring the distributions of the electric field and the curl of the the monopole current. The derivation of the Gauss law on the lattice demonstrates how to define the electric field strength as an $8$-vector in colour space (\ref{fields1}). To determine it as a gauge invariant quantity we have to evaluate - according to (\ref{gauss}) - the correlation function
\begin{equation}
\label{Ecorrel}
\langle E_i^{x_+}(x) \rangle_{Q\bar{Q}} \;\;\; = \;\;\; \frac{\langle \mbox{Tr}\left( U(x_+,x)E_i^x(x)U^\dagger(x_+,x)L(x_+)\right) \mbox{Tr} L^\ast(\vec{x}_-) \rangle}{\langle \mbox{Tr}L(\vec{x}_+) \mbox{Tr}L^\ast(\vec{x}_-)\rangle}
\end{equation}
with $U(x_+,x)$ being the Schwinger line connecting the field strength $E_i^x(x)$ with the Polyakov loop $L(x_+)$ (Fig.\ref{BildE}). This correlation is very similar to the one proposed in \cite{stefan} for the electric field. To measure the distribution of the curl of the monopole current we generalize (\ref{Ecorrel}) and replace the electric field strength by the curl defined in (\ref{rot}). Because of local gauge invariance of QCD in colour space we have to refer both operators in (\ref{Ecorrel}), Polyakov loop and field strength, to the same lattice site. This is guaranteed by the Schwinger line in (\ref{Ecorrel}). But there is no unique choice of Schwinger line connecting the two operators. Since we try to verify the validity of the dual London equation what means comparing two quantities with each other, this problem has no consequences to our intention, if the electric field strength and the curl of the monopole current are transported on the same path and therefore measured in the same local coordinate system.
There is another property of the correlation (\ref{Ecorrel}) which should be mentioned. Charge and anticharge do not play a symmetric role as one would naively expect, because the electric field can only be connected with either the charge or the anticharge. Therefore we should not expect a result which is symmetric with respect to the two charges.   

Now we consider the numerical results which we obtain from evaluating the correlation function (\ref{Ecorrel}) for the electric field and the curl of the monopole current. We used $50.000$ gauge field configurations and varied the distance between the charges from $1a$ to $4a$.
\begin{figure}
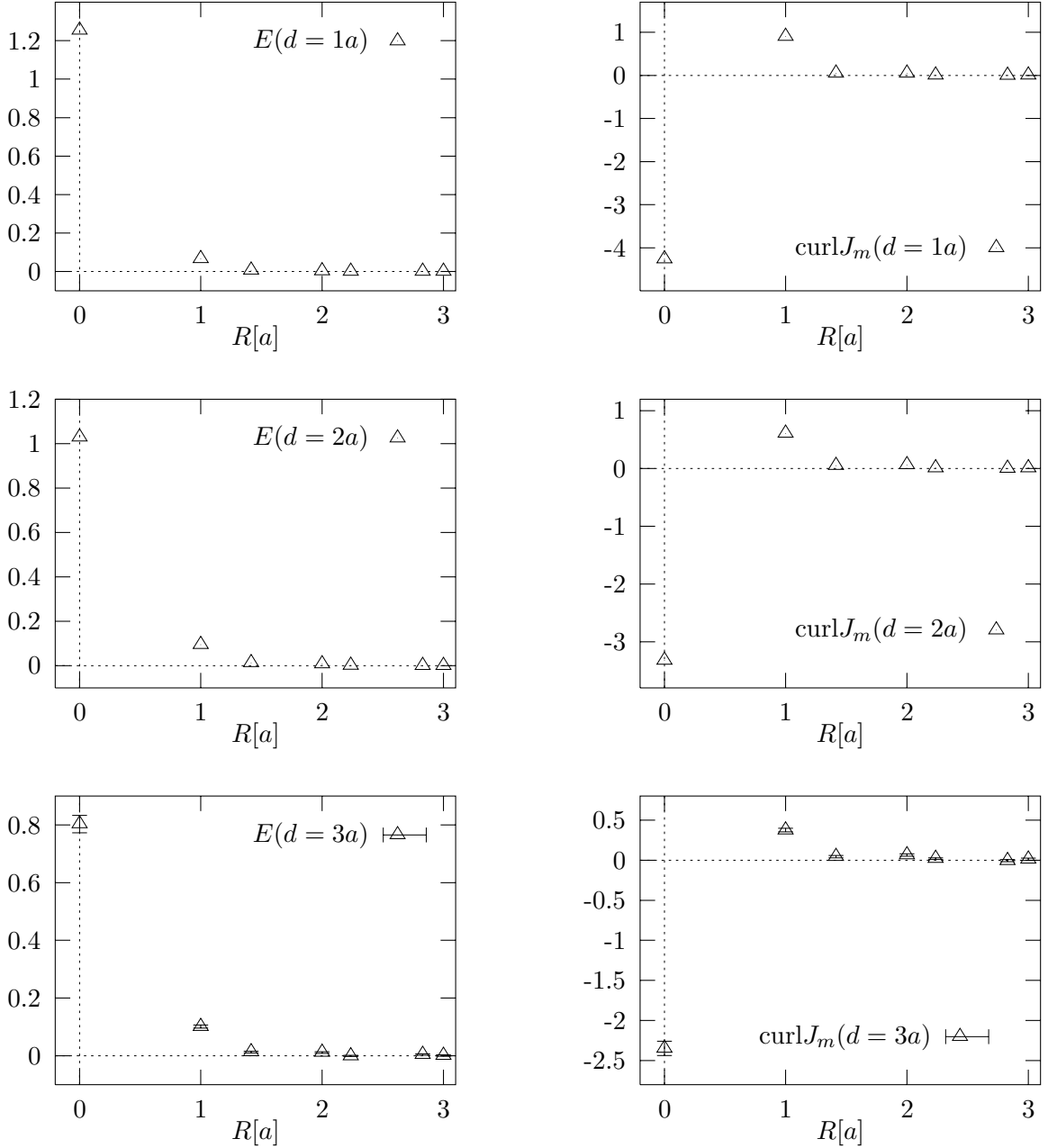

\vspace{-1cm}
\centerline{\input{fig7a.tex}\input{fig7b.tex}}
\vspace{0.5cm}
\centerline{\input{fig7c.tex}\input{fig7d.tex}}
\vspace{0.5cm}
\centerline{\input{fig7e.tex}\input{fig7f.tex}}
\vspace{0.3cm}
\caption{\label{bild1}Transverse profile of the electric field strength (left column) and the curl of the monopole current (right column) in the plane perpendicular to the $Q\bar{Q}$-axis for charge distances $d=1a$, $d=2a$ and $d=3a$. The electric field strength decreases quickly with increasing radial distance from the $Q\bar{Q}$-axis. This indicates that the flux tube is mainly concentrated on the link(s) connecting the charges. The absolute value of the curl shows the same behaviour as the electric field strength, only on the link(s) connecting the charges they have a different sign because of the fluxoid. The errorbars shown indicate the jack knife error, for $d=1a$ and $d=2a$ they have been omitted, because they are smaller than the plotting symbols.}
\end{figure}
\begin{figure}
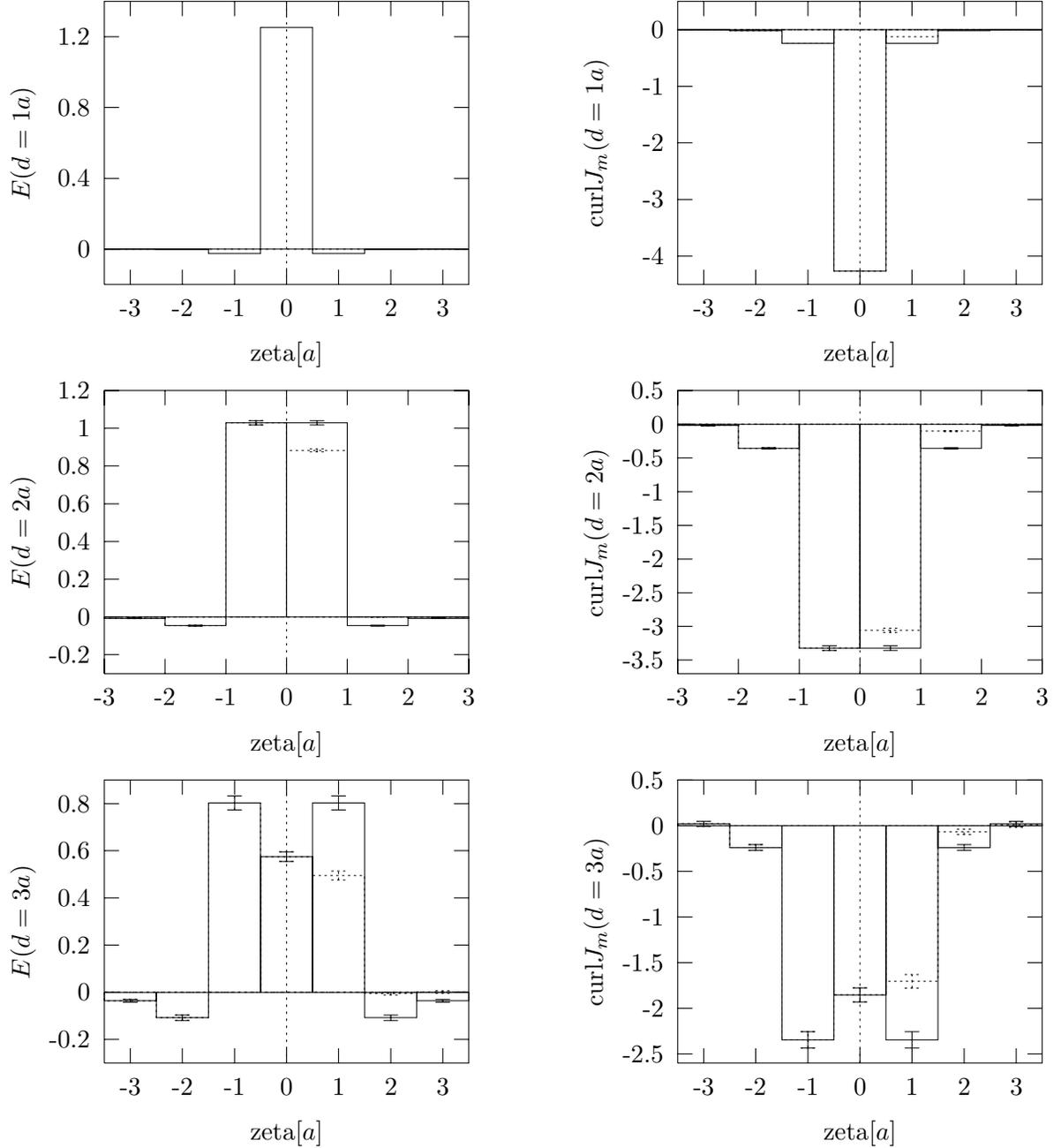

\vspace{-1cm}
\centerline{\input{fig8a.tex}\input{fig8b.tex}}
\vspace{0.5cm}
\centerline{
\setlength{\unitlength}{0.1bp}
\special{!
/gnudict 40 dict def
gnudict begin
/Color false def
/Solid false def
/gnulinewidth 2.000 def
/vshift -33 def
/dl {10 mul} def
/hpt 31.5 def
/vpt 31.5 def
/M {moveto} bind def
/L {lineto} bind def
/R {rmoveto} bind def
/V {rlineto} bind def
/vpt2 vpt 2 mul def
/hpt2 hpt 2 mul def
/Lshow { currentpoint stroke M
  0 vshift R show } def
/Rshow { currentpoint stroke M
  dup stringwidth pop neg vshift R show } def
/Cshow { currentpoint stroke M
  dup stringwidth pop -2 div vshift R show } def
/DL { Color {setrgbcolor Solid {pop []} if 0 setdash }
 {pop pop pop Solid {pop []} if 0 setdash} ifelse } def
/BL { stroke gnulinewidth 2 mul setlinewidth } def
/AL { stroke gnulinewidth 2 div setlinewidth } def
/PL { stroke gnulinewidth setlinewidth } def
/LTb { BL [] 0 0 0 DL } def
/LTa { AL [1 dl 2 dl] 0 setdash 0 0 0 setrgbcolor } def
/LT0 { PL [] 0 1 0 DL } def
/LT1 { PL [4 dl 2 dl] 0 0 1 DL } def
/LT2 { PL [2 dl 3 dl] 1 0 0 DL } def
/LT3 { PL [1 dl 1.5 dl] 1 0 1 DL } def
/LT4 { PL [5 dl 2 dl 1 dl 2 dl] 0 1 1 DL } def
/LT5 { PL [4 dl 3 dl 1 dl 3 dl] 1 1 0 DL } def
/LT6 { PL [2 dl 2 dl 2 dl 4 dl] 0 0 0 DL } def
/LT7 { PL [2 dl 2 dl 2 dl 2 dl 2 dl 4 dl] 1 0.3 0 DL } def
/LT8 { PL [2 dl 2 dl 2 dl 2 dl 2 dl 2 dl 2 dl 4 dl] 0.5 0.5 0.5 DL } def
/P { stroke [] 0 setdash
  currentlinewidth 2 div sub M
  0 currentlinewidth V stroke } def
/D { stroke [] 0 setdash 2 copy vpt add M
  hpt neg vpt neg V hpt vpt neg V
  hpt vpt V hpt neg vpt V closepath stroke
  P } def
/A { stroke [] 0 setdash vpt sub M 0 vpt2 V
  currentpoint stroke M
  hpt neg vpt neg R hpt2 0 V stroke
  } def
/B { stroke [] 0 setdash 2 copy exch hpt sub exch vpt add M
  0 vpt2 neg V hpt2 0 V 0 vpt2 V
  hpt2 neg 0 V closepath stroke
  P } def
/C { stroke [] 0 setdash exch hpt sub exch vpt add M
  hpt2 vpt2 neg V currentpoint stroke M
  hpt2 neg 0 R hpt2 vpt2 V stroke } def
/T { stroke [] 0 setdash 2 copy vpt 1.12 mul add M
  hpt neg vpt -1.62 mul V
  hpt 2 mul 0 V
  hpt neg vpt 1.62 mul V closepath stroke
  P  } def
/S { 2 copy A C} def
end
}
\begin{picture}(2339,1511)(0,0)
\special{"
gnudict begin
gsave
50 50 translate
0.100 0.100 scale
0 setgray
/Helvetica findfont 100 scalefont setfont
newpath
-500.000000 -500.000000 translate
LTa
600 493 M
1556 0 V
1378 251 M
0 1209 V
LTb
600 332 M
63 0 V
1493 0 R
-63 0 V
600 493 M
63 0 V
1493 0 R
-63 0 V
600 654 M
63 0 V
1493 0 R
-63 0 V
600 815 M
63 0 V
1493 0 R
-63 0 V
600 976 M
63 0 V
1493 0 R
-63 0 V
600 1138 M
63 0 V
1493 0 R
-63 0 V
600 1299 M
63 0 V
1493 0 R
-63 0 V
600 1460 M
63 0 V
1493 0 R
-63 0 V
600 251 M
0 63 V
0 1146 R
0 -63 V
859 251 M
0 63 V
0 1146 R
0 -63 V
1119 251 M
0 63 V
0 1146 R
0 -63 V
1378 251 M
0 63 V
0 1146 R
0 -63 V
1637 251 M
0 63 V
0 1146 R
0 -63 V
1897 251 M
0 63 V
0 1146 R
0 -63 V
2156 251 M
0 63 V
0 1146 R
0 -63 V
600 251 M
1556 0 V
0 1209 V
-1556 0 V
600 251 L
LT0
LT3
730 486 M
0 2 V
-31 -2 R
62 0 V
-62 2 R
62 0 V
989 454 M
0 4 V
-31 -4 R
62 0 V
-62 4 R
62 0 V
228 855 R
0 18 V
-31 -18 R
62 0 V
-62 18 R
62 0 V
229 -133 R
0 13 V
-31 -13 R
62 0 V
-62 13 R
62 0 V
1767 492 M
0 2 V
-31 -2 R
62 0 V
-62 2 R
62 0 V
228 -2 R
0 2 V
-31 -2 R
62 0 V
-62 2 R
62 0 V
600 493 M
0 -6 V
259 0 V
0 6 V
-259 0 V
259 0 R
0 -37 V
260 0 V
0 37 V
-260 0 V
260 0 R
0 829 V
259 0 V
0 -829 V
-259 0 V
259 0 R
0 711 V
259 0 V
0 -711 V
-259 0 V
259 0 R
260 0 V
-260 0 V
260 0 R
259 0 V
-259 0 V
LT0
730 486 M
0 2 V
-31 -2 R
62 0 V
-62 2 R
62 0 V
989 454 M
0 4 V
-31 -4 R
62 0 V
-62 4 R
62 0 V
228 855 R
0 18 V
-31 -18 R
62 0 V
-62 18 R
62 0 V
229 -18 R
0 18 V
-31 -18 R
62 0 V
-62 18 R
62 0 V
1767 454 M
0 4 V
-31 -4 R
62 0 V
-62 4 R
62 0 V
228 28 R
0 2 V
-31 -2 R
62 0 V
-62 2 R
62 0 V
600 493 M
0 -6 V
259 0 V
0 6 V
-259 0 V
259 0 R
0 -37 V
260 0 V
0 37 V
-260 0 V
260 0 R
0 829 V
259 0 V
0 -829 V
-259 0 V
259 0 R
0 829 V
259 0 V
0 -829 V
-259 0 V
259 0 R
0 -37 V
260 0 V
0 37 V
-260 0 V
260 0 R
0 -6 V
259 0 V
0 6 V
-259 0 V
stroke
grestore
end
showpage
}
\put(1378,-49){\makebox(0,0){$\mbox{zeta}[a]$}}
\put(280,855){%
\special{ps: gsave currentpoint currentpoint translate
270 rotate neg exch neg exch translate}%
\makebox(0,0)[b]{\shortstack{$E(d=2a)$}}%
\special{ps: currentpoint grestore moveto}%
}
\put(2156,151){\makebox(0,0){3}}
\put(1897,151){\makebox(0,0){2}}
\put(1637,151){\makebox(0,0){1}}
\put(1378,151){\makebox(0,0){0}}
\put(1119,151){\makebox(0,0){-1}}
\put(859,151){\makebox(0,0){-2}}
\put(600,151){\makebox(0,0){-3}}
\put(540,1460){\makebox(0,0)[r]{1.2}}
\put(540,1299){\makebox(0,0)[r]{1}}
\put(540,1138){\makebox(0,0)[r]{0.8}}
\put(540,976){\makebox(0,0)[r]{0.6}}
\put(540,815){\makebox(0,0)[r]{0.4}}
\put(540,654){\makebox(0,0)[r]{0.2}}
\put(540,493){\makebox(0,0)[r]{0}}
\put(540,332){\makebox(0,0)[r]{-0.2}}
\end{picture}
\setlength{\unitlength}{0.1bp}
\special{!
/gnudict 40 dict def
gnudict begin
/Color false def
/Solid false def
/gnulinewidth 2.000 def
/vshift -33 def
/dl {10 mul} def
/hpt 31.5 def
/vpt 31.5 def
/M {moveto} bind def
/L {lineto} bind def
/R {rmoveto} bind def
/V {rlineto} bind def
/vpt2 vpt 2 mul def
/hpt2 hpt 2 mul def
/Lshow { currentpoint stroke M
  0 vshift R show } def
/Rshow { currentpoint stroke M
  dup stringwidth pop neg vshift R show } def
/Cshow { currentpoint stroke M
  dup stringwidth pop -2 div vshift R show } def
/DL { Color {setrgbcolor Solid {pop []} if 0 setdash }
 {pop pop pop Solid {pop []} if 0 setdash} ifelse } def
/BL { stroke gnulinewidth 2 mul setlinewidth } def
/AL { stroke gnulinewidth 2 div setlinewidth } def
/PL { stroke gnulinewidth setlinewidth } def
/LTb { BL [] 0 0 0 DL } def
/LTa { AL [1 dl 2 dl] 0 setdash 0 0 0 setrgbcolor } def
/LT0 { PL [] 0 1 0 DL } def
/LT1 { PL [4 dl 2 dl] 0 0 1 DL } def
/LT2 { PL [2 dl 3 dl] 1 0 0 DL } def
/LT3 { PL [1 dl 1.5 dl] 1 0 1 DL } def
/LT4 { PL [5 dl 2 dl 1 dl 2 dl] 0 1 1 DL } def
/LT5 { PL [4 dl 3 dl 1 dl 3 dl] 1 1 0 DL } def
/LT6 { PL [2 dl 2 dl 2 dl 4 dl] 0 0 0 DL } def
/LT7 { PL [2 dl 2 dl 2 dl 2 dl 2 dl 4 dl] 1 0.3 0 DL } def
/LT8 { PL [2 dl 2 dl 2 dl 2 dl 2 dl 2 dl 2 dl 4 dl] 0.5 0.5 0.5 DL } def
/P { stroke [] 0 setdash
  currentlinewidth 2 div sub M
  0 currentlinewidth V stroke } def
/D { stroke [] 0 setdash 2 copy vpt add M
  hpt neg vpt neg V hpt vpt neg V
  hpt vpt V hpt neg vpt V closepath stroke
  P } def
/A { stroke [] 0 setdash vpt sub M 0 vpt2 V
  currentpoint stroke M
  hpt neg vpt neg R hpt2 0 V stroke
  } def
/B { stroke [] 0 setdash 2 copy exch hpt sub exch vpt add M
  0 vpt2 neg V hpt2 0 V 0 vpt2 V
  hpt2 neg 0 V closepath stroke
  P } def
/C { stroke [] 0 setdash exch hpt sub exch vpt add M
  hpt2 vpt2 neg V currentpoint stroke M
  hpt2 neg 0 R hpt2 vpt2 V stroke } def
/T { stroke [] 0 setdash 2 copy vpt 1.12 mul add M
  hpt neg vpt -1.62 mul V
  hpt 2 mul 0 V
  hpt neg vpt 1.62 mul V closepath stroke
  P  } def
/S { 2 copy A C} def
end
}
\begin{picture}(2339,1511)(0,0)
\special{"
gnudict begin
gsave
50 50 translate
0.100 0.100 scale
0 setgray
/Helvetica findfont 100 scalefont setfont
newpath
-500.000000 -500.000000 translate
LTa
600 1316 M
1556 0 V
1378 251 M
0 1209 V
LTb
600 309 M
63 0 V
1493 0 R
-63 0 V
600 453 M
63 0 V
1493 0 R
-63 0 V
600 596 M
63 0 V
1493 0 R
-63 0 V
600 740 M
63 0 V
1493 0 R
-63 0 V
600 884 M
63 0 V
1493 0 R
-63 0 V
600 1028 M
63 0 V
1493 0 R
-63 0 V
600 1172 M
63 0 V
1493 0 R
-63 0 V
600 1316 M
63 0 V
1493 0 R
-63 0 V
600 1460 M
63 0 V
1493 0 R
-63 0 V
600 251 M
0 63 V
0 1146 R
0 -63 V
859 251 M
0 63 V
0 1146 R
0 -63 V
1119 251 M
0 63 V
0 1146 R
0 -63 V
1378 251 M
0 63 V
0 1146 R
0 -63 V
1637 251 M
0 63 V
0 1146 R
0 -63 V
1897 251 M
0 63 V
0 1146 R
0 -63 V
2156 251 M
0 63 V
0 1146 R
0 -63 V
600 251 M
1556 0 V
0 1209 V
-1556 0 V
600 251 L
LT0
LT3
730 1309 M
0 4 V
-31 -4 R
62 0 V
-62 4 R
62 0 V
989 1210 M
0 6 V
-31 -6 R
62 0 V
-62 6 R
62 0 V
1248 350 M
0 20 V
-31 -20 R
62 0 V
-62 20 R
62 0 V
229 57 R
0 18 V
-31 -18 R
62 0 V
-62 18 R
62 0 V
228 839 R
0 5 V
-31 -5 R
62 0 V
-62 5 R
62 0 V
228 21 R
0 5 V
-31 -5 R
62 0 V
-62 5 R
62 0 V
-1457 1 R
0 -5 V
259 0 V
0 5 V
-259 0 V
259 0 R
0 -103 V
260 0 V
0 103 V
-260 0 V
260 0 R
0 -956 V
259 0 V
0 956 V
-259 0 V
259 0 R
0 -880 V
259 0 V
0 880 V
-259 0 V
259 0 R
0 -29 V
260 0 V
0 29 V
-260 0 V
260 0 R
0 -4 V
259 0 V
0 4 V
-259 0 V
LT0
730 1309 M
0 4 V
-31 -4 R
62 0 V
-62 4 R
62 0 V
989 1210 M
0 6 V
-31 -6 R
62 0 V
-62 6 R
62 0 V
1248 350 M
0 20 V
-31 -20 R
62 0 V
-62 20 R
62 0 V
229 -20 R
0 20 V
-31 -20 R
62 0 V
-62 20 R
62 0 V
228 840 R
0 6 V
-31 -6 R
62 0 V
-62 6 R
62 0 V
228 93 R
0 4 V
-31 -4 R
62 0 V
-62 4 R
62 0 V
-1457 3 R
0 -5 V
259 0 V
0 5 V
-259 0 V
259 0 R
0 -103 V
260 0 V
0 103 V
-260 0 V
260 0 R
0 -956 V
259 0 V
0 956 V
-259 0 V
259 0 R
0 -956 V
259 0 V
0 956 V
-259 0 V
259 0 R
0 -103 V
260 0 V
0 103 V
-260 0 V
260 0 R
0 -5 V
259 0 V
0 5 V
-259 0 V
stroke
grestore
end
showpage
}
\put(1378,-49){\makebox(0,0){$\mbox{zeta}[a]$}}
\put(280,855){%
\special{ps: gsave currentpoint currentpoint translate
270 rotate neg exch neg exch translate}%
\makebox(0,0)[b]{\shortstack{$\mbox{curl}J_m(d=2a)$}}%
\special{ps: currentpoint grestore moveto}%
}
\put(2156,151){\makebox(0,0){3}}
\put(1897,151){\makebox(0,0){2}}
\put(1637,151){\makebox(0,0){1}}
\put(1378,151){\makebox(0,0){0}}
\put(1119,151){\makebox(0,0){-1}}
\put(859,151){\makebox(0,0){-2}}
\put(600,151){\makebox(0,0){-3}}
\put(540,1460){\makebox(0,0)[r]{0.5}}
\put(540,1316){\makebox(0,0)[r]{0}}
\put(540,1172){\makebox(0,0)[r]{-0.5}}
\put(540,1028){\makebox(0,0)[r]{-1}}
\put(540,884){\makebox(0,0)[r]{-1.5}}
\put(540,740){\makebox(0,0)[r]{-2}}
\put(540,596){\makebox(0,0)[r]{-2.5}}
\put(540,453){\makebox(0,0)[r]{-3}}
\put(540,309){\makebox(0,0)[r]{-3.5}}
\end{picture}}
\vspace{0.5cm}
\centerline{
\setlength{\unitlength}{0.1bp}
\special{!
/gnudict 40 dict def
gnudict begin
/Color false def
/Solid false def
/gnulinewidth 2.000 def
/vshift -33 def
/dl {10 mul} def
/hpt 31.5 def
/vpt 31.5 def
/M {moveto} bind def
/L {lineto} bind def
/R {rmoveto} bind def
/V {rlineto} bind def
/vpt2 vpt 2 mul def
/hpt2 hpt 2 mul def
/Lshow { currentpoint stroke M
  0 vshift R show } def
/Rshow { currentpoint stroke M
  dup stringwidth pop neg vshift R show } def
/Cshow { currentpoint stroke M
  dup stringwidth pop -2 div vshift R show } def
/DL { Color {setrgbcolor Solid {pop []} if 0 setdash }
 {pop pop pop Solid {pop []} if 0 setdash} ifelse } def
/BL { stroke gnulinewidth 2 mul setlinewidth } def
/AL { stroke gnulinewidth 2 div setlinewidth } def
/PL { stroke gnulinewidth setlinewidth } def
/LTb { BL [] 0 0 0 DL } def
/LTa { AL [1 dl 2 dl] 0 setdash 0 0 0 setrgbcolor } def
/LT0 { PL [] 0 1 0 DL } def
/LT1 { PL [4 dl 2 dl] 0 0 1 DL } def
/LT2 { PL [2 dl 3 dl] 1 0 0 DL } def
/LT3 { PL [1 dl 1.5 dl] 1 0 1 DL } def
/LT4 { PL [5 dl 2 dl 1 dl 2 dl] 0 1 1 DL } def
/LT5 { PL [4 dl 3 dl 1 dl 3 dl] 1 1 0 DL } def
/LT6 { PL [2 dl 2 dl 2 dl 4 dl] 0 0 0 DL } def
/LT7 { PL [2 dl 2 dl 2 dl 2 dl 2 dl 4 dl] 1 0.3 0 DL } def
/LT8 { PL [2 dl 2 dl 2 dl 2 dl 2 dl 2 dl 2 dl 4 dl] 0.5 0.5 0.5 DL } def
/P { stroke [] 0 setdash
  currentlinewidth 2 div sub M
  0 currentlinewidth V stroke } def
/D { stroke [] 0 setdash 2 copy vpt add M
  hpt neg vpt neg V hpt vpt neg V
  hpt vpt V hpt neg vpt V closepath stroke
  P } def
/A { stroke [] 0 setdash vpt sub M 0 vpt2 V
  currentpoint stroke M
  hpt neg vpt neg R hpt2 0 V stroke
  } def
/B { stroke [] 0 setdash 2 copy exch hpt sub exch vpt add M
  0 vpt2 neg V hpt2 0 V 0 vpt2 V
  hpt2 neg 0 V closepath stroke
  P } def
/C { stroke [] 0 setdash exch hpt sub exch vpt add M
  hpt2 vpt2 neg V currentpoint stroke M
  hpt2 neg 0 R hpt2 vpt2 V stroke } def
/T { stroke [] 0 setdash 2 copy vpt 1.12 mul add M
  hpt neg vpt -1.62 mul V
  hpt 2 mul 0 V
  hpt neg vpt 1.62 mul V closepath stroke
  P  } def
/S { 2 copy A C} def
end
}
\begin{picture}(2339,1511)(0,0)
\special{"
gnudict begin
gsave
50 50 translate
0.100 0.100 scale
0 setgray
/Helvetica findfont 100 scalefont setfont
newpath
-500.000000 -500.000000 translate
LTa
600 553 M
1556 0 V
1378 251 M
0 1209 V
LTb
600 352 M
63 0 V
1493 0 R
-63 0 V
600 553 M
63 0 V
1493 0 R
-63 0 V
600 755 M
63 0 V
1493 0 R
-63 0 V
600 956 M
63 0 V
1493 0 R
-63 0 V
600 1158 M
63 0 V
1493 0 R
-63 0 V
600 1359 M
63 0 V
1493 0 R
-63 0 V
711 251 M
0 63 V
0 1146 R
0 -63 V
933 251 M
0 63 V
0 1146 R
0 -63 V
1156 251 M
0 63 V
0 1146 R
0 -63 V
1378 251 M
0 63 V
0 1146 R
0 -63 V
1600 251 M
0 63 V
0 1146 R
0 -63 V
1823 251 M
0 63 V
0 1146 R
0 -63 V
2045 251 M
0 63 V
0 1146 R
0 -63 V
600 251 M
1556 0 V
0 1209 V
-1556 0 V
600 251 L
LT0
LT3
711 511 M
0 12 V
680 511 M
62 0 V
-62 12 R
62 0 V
933 433 M
0 23 V
902 433 M
62 0 V
-62 23 R
62 0 V
192 876 R
0 60 V
-31 -60 R
62 0 V
-62 60 R
62 0 V
191 -280 R
0 41 V
-31 -41 R
62 0 V
-62 41 R
62 0 V
191 -120 R
0 38 V
-31 -38 R
62 0 V
-62 38 R
62 0 V
1823 542 M
0 11 V
-31 -11 R
62 0 V
-62 11 R
62 0 V
191 -4 R
0 11 V
-31 -11 R
62 0 V
-62 11 R
62 0 V
600 553 M
0 -36 V
222 0 V
0 36 V
-222 0 V
222 0 R
0 -108 V
223 0 V
0 108 V
-223 0 V
223 0 R
0 809 V
222 0 V
0 -809 V
-222 0 V
222 0 R
0 579 V
222 0 V
0 -579 V
-222 0 V
222 0 R
0 499 V
222 0 V
0 -499 V
-222 0 V
222 0 R
0 -5 V
223 0 V
0 5 V
-223 0 V
223 0 R
0 1 V
222 0 V
0 -1 V
-222 0 V
LT0
711 511 M
0 12 V
680 511 M
62 0 V
-62 12 R
62 0 V
933 433 M
0 23 V
902 433 M
62 0 V
-62 23 R
62 0 V
192 876 R
0 60 V
-31 -60 R
62 0 V
-62 60 R
62 0 V
191 -280 R
0 41 V
-31 -41 R
62 0 V
-62 41 R
62 0 V
191 179 R
0 60 V
-31 -60 R
62 0 V
-62 60 R
62 0 V
1823 433 M
0 23 V
-31 -23 R
62 0 V
-62 23 R
62 0 V
191 55 R
0 12 V
-31 -12 R
62 0 V
-62 12 R
62 0 V
600 553 M
0 -36 V
222 0 V
0 36 V
-222 0 V
222 0 R
0 -108 V
223 0 V
0 108 V
-223 0 V
223 0 R
0 809 V
222 0 V
0 -809 V
-222 0 V
222 0 R
0 579 V
222 0 V
0 -579 V
-222 0 V
222 0 R
0 809 V
222 0 V
0 -809 V
-222 0 V
222 0 R
0 -108 V
223 0 V
0 108 V
-223 0 V
223 0 R
0 -36 V
222 0 V
0 36 V
-222 0 V
stroke
grestore
end
showpage
}
\put(1378,-49){\makebox(0,0){$\mbox{zeta}[a]$}}
\put(280,855){%
\special{ps: gsave currentpoint currentpoint translate
270 rotate neg exch neg exch translate}%
\makebox(0,0)[b]{\shortstack{$E(d=3a)$}}%
\special{ps: currentpoint grestore moveto}%
}
\put(2045,151){\makebox(0,0){3}}
\put(1823,151){\makebox(0,0){2}}
\put(1600,151){\makebox(0,0){1}}
\put(1378,151){\makebox(0,0){0}}
\put(1156,151){\makebox(0,0){-1}}
\put(933,151){\makebox(0,0){-2}}
\put(711,151){\makebox(0,0){-3}}
\put(540,1359){\makebox(0,0)[r]{0.8}}
\put(540,1158){\makebox(0,0)[r]{0.6}}
\put(540,956){\makebox(0,0)[r]{0.4}}
\put(540,755){\makebox(0,0)[r]{0.2}}
\put(540,553){\makebox(0,0)[r]{0}}
\put(540,352){\makebox(0,0)[r]{-0.2}}
\end{picture}
\setlength{\unitlength}{0.1bp}
\special{!
/gnudict 40 dict def
gnudict begin
/Color false def
/Solid false def
/gnulinewidth 2.000 def
/vshift -33 def
/dl {10 mul} def
/hpt 31.5 def
/vpt 31.5 def
/M {moveto} bind def
/L {lineto} bind def
/R {rmoveto} bind def
/V {rlineto} bind def
/vpt2 vpt 2 mul def
/hpt2 hpt 2 mul def
/Lshow { currentpoint stroke M
  0 vshift R show } def
/Rshow { currentpoint stroke M
  dup stringwidth pop neg vshift R show } def
/Cshow { currentpoint stroke M
  dup stringwidth pop -2 div vshift R show } def
/DL { Color {setrgbcolor Solid {pop []} if 0 setdash }
 {pop pop pop Solid {pop []} if 0 setdash} ifelse } def
/BL { stroke gnulinewidth 2 mul setlinewidth } def
/AL { stroke gnulinewidth 2 div setlinewidth } def
/PL { stroke gnulinewidth setlinewidth } def
/LTb { BL [] 0 0 0 DL } def
/LTa { AL [1 dl 2 dl] 0 setdash 0 0 0 setrgbcolor } def
/LT0 { PL [] 0 1 0 DL } def
/LT1 { PL [4 dl 2 dl] 0 0 1 DL } def
/LT2 { PL [2 dl 3 dl] 1 0 0 DL } def
/LT3 { PL [1 dl 1.5 dl] 1 0 1 DL } def
/LT4 { PL [5 dl 2 dl 1 dl 2 dl] 0 1 1 DL } def
/LT5 { PL [4 dl 3 dl 1 dl 3 dl] 1 1 0 DL } def
/LT6 { PL [2 dl 2 dl 2 dl 4 dl] 0 0 0 DL } def
/LT7 { PL [2 dl 2 dl 2 dl 2 dl 2 dl 4 dl] 1 0.3 0 DL } def
/LT8 { PL [2 dl 2 dl 2 dl 2 dl 2 dl 2 dl 2 dl 4 dl] 0.5 0.5 0.5 DL } def
/P { stroke [] 0 setdash
  currentlinewidth 2 div sub M
  0 currentlinewidth V stroke } def
/D { stroke [] 0 setdash 2 copy vpt add M
  hpt neg vpt neg V hpt vpt neg V
  hpt vpt V hpt neg vpt V closepath stroke
  P } def
/A { stroke [] 0 setdash vpt sub M 0 vpt2 V
  currentpoint stroke M
  hpt neg vpt neg R hpt2 0 V stroke
  } def
/B { stroke [] 0 setdash 2 copy exch hpt sub exch vpt add M
  0 vpt2 neg V hpt2 0 V 0 vpt2 V
  hpt2 neg 0 V closepath stroke
  P } def
/C { stroke [] 0 setdash exch hpt sub exch vpt add M
  hpt2 vpt2 neg V currentpoint stroke M
  hpt2 neg 0 R hpt2 vpt2 V stroke } def
/T { stroke [] 0 setdash 2 copy vpt 1.12 mul add M
  hpt neg vpt -1.62 mul V
  hpt 2 mul 0 V
  hpt neg vpt 1.62 mul V closepath stroke
  P  } def
/S { 2 copy A C} def
end
}
\begin{picture}(2339,1511)(0,0)
\special{"
gnudict begin
gsave
50 50 translate
0.100 0.100 scale
0 setgray
/Helvetica findfont 100 scalefont setfont
newpath
-500.000000 -500.000000 translate
LTa
600 1265 M
1556 0 V
1378 251 M
0 1209 V
LTb
600 290 M
63 0 V
1493 0 R
-63 0 V
600 485 M
63 0 V
1493 0 R
-63 0 V
600 680 M
63 0 V
1493 0 R
-63 0 V
600 875 M
63 0 V
1493 0 R
-63 0 V
600 1070 M
63 0 V
1493 0 R
-63 0 V
600 1265 M
63 0 V
1493 0 R
-63 0 V
600 1460 M
63 0 V
1493 0 R
-63 0 V
711 251 M
0 63 V
0 1146 R
0 -63 V
933 251 M
0 63 V
0 1146 R
0 -63 V
1156 251 M
0 63 V
0 1146 R
0 -63 V
1378 251 M
0 63 V
0 1146 R
0 -63 V
1600 251 M
0 63 V
0 1146 R
0 -63 V
1823 251 M
0 63 V
0 1146 R
0 -63 V
2045 251 M
0 63 V
0 1146 R
0 -63 V
600 251 M
1556 0 V
0 1209 V
-1556 0 V
600 251 L
LT0
LT3
711 1262 M
0 21 V
-31 -21 R
62 0 V
-62 21 R
62 0 V
933 1159 M
0 25 V
-31 -25 R
62 0 V
-62 25 R
62 0 V
1156 315 M
0 70 V
-31 -70 R
62 0 V
-62 70 R
62 0 V
191 127 R
0 60 V
-31 -60 R
62 0 V
-62 60 R
62 0 V
191 -1 R
0 58 V
-31 -58 R
62 0 V
-62 58 R
62 0 V
192 598 R
0 23 V
-31 -23 R
62 0 V
-62 23 R
62 0 V
191 8 R
0 23 V
-31 -23 R
62 0 V
-62 23 R
62 0 V
600 1265 M
0 8 V
222 0 V
0 -8 V
-222 0 V
222 0 R
0 -94 V
223 0 V
0 94 V
-223 0 V
223 0 R
0 -915 V
222 0 V
0 915 V
-222 0 V
222 0 R
0 -723 V
222 0 V
0 723 V
-222 0 V
222 0 R
0 -665 V
222 0 V
0 665 V
-222 0 V
222 0 R
0 -26 V
223 0 V
0 26 V
-223 0 V
223 0 R
0 5 V
222 0 V
0 -5 V
-222 0 V
LT0
711 1262 M
0 21 V
-31 -21 R
62 0 V
-62 21 R
62 0 V
933 1159 M
0 25 V
-31 -25 R
62 0 V
-62 25 R
62 0 V
1156 315 M
0 70 V
-31 -70 R
62 0 V
-62 70 R
62 0 V
191 127 R
0 60 V
-31 -60 R
62 0 V
-62 60 R
62 0 V
1600 315 M
0 70 V
-31 -70 R
62 0 V
-62 70 R
62 0 V
192 774 R
0 25 V
-31 -25 R
62 0 V
-62 25 R
62 0 V
191 78 R
0 21 V
-31 -21 R
62 0 V
-62 21 R
62 0 V
600 1265 M
0 8 V
222 0 V
0 -8 V
-222 0 V
222 0 R
0 -94 V
223 0 V
0 94 V
-223 0 V
223 0 R
0 -915 V
222 0 V
0 915 V
-222 0 V
222 0 R
0 -723 V
222 0 V
0 723 V
-222 0 V
222 0 R
0 -915 V
222 0 V
0 915 V
-222 0 V
222 0 R
0 -94 V
223 0 V
0 94 V
-223 0 V
223 0 R
0 8 V
222 0 V
0 -8 V
-222 0 V
stroke
grestore
end
showpage
}
\put(1378,-49){\makebox(0,0){$\mbox{zeta}[a]$}}
\put(280,855){%
\special{ps: gsave currentpoint currentpoint translate
270 rotate neg exch neg exch translate}%
\makebox(0,0)[b]{\shortstack{$\mbox{curl}J_m(d=3a)$}}%
\special{ps: currentpoint grestore moveto}%
}
\put(2045,151){\makebox(0,0){3}}
\put(1823,151){\makebox(0,0){2}}
\put(1600,151){\makebox(0,0){1}}
\put(1378,151){\makebox(0,0){0}}
\put(1156,151){\makebox(0,0){-1}}
\put(933,151){\makebox(0,0){-2}}
\put(711,151){\makebox(0,0){-3}}
\put(540,1460){\makebox(0,0)[r]{0.5}}
\put(540,1265){\makebox(0,0)[r]{0}}
\put(540,1070){\makebox(0,0)[r]{-0.5}}
\put(540,875){\makebox(0,0)[r]{-1}}
\put(540,680){\makebox(0,0)[r]{-1.5}}
\put(540,485){\makebox(0,0)[r]{-2}}
\put(540,290){\makebox(0,0)[r]{-2.5}}
\end{picture}}
\vspace{0.3cm}
\caption{\label{bild2}Longitudinal profile of the electric field strength (left column) and the curl of the monopole current (right column) on the $Q\bar{Q}$-axis for charge distances $d=1a$, $d=2a$ and $d=3a$. For $d=1a$ the charges are at $\mbox{zeta}=\pm0.5a$, for $d=2a$ at $\mbox{zeta}=\pm1a$ and for $d=3a$ at $\mbox{zeta}=\pm1.5a$. For all three charge distances the electric flux as well as the curl of the monopole current is mainly concentrated on the link(s) connecting the two charges. As described in the text the full (dotted) line corresponds to a symmetrized (non symmetrized) ``measurement''. The error bars shown indicate the jack knife error.}
\end{figure}
In Fig.\ref{bild1} we show a transverse profile of the electric field strength (left column of Fig.\ref{bild1}) and the curl of the monopole current (right column of Fig.\ref{bild1}) in the plane perpendicular to the axis connecting the charges for charge distances $d=1a$, $d=2a$ and $d=3a$. The electric field strength decreases quickly with increasing radial distance from the $Q\bar{Q}$-axis what characterizes a thin flux tube between the charges. The absolute value of the curl of the monopole current shows the same behaviour as the electric field, only on the $Q\bar{Q}$-axis the curl is reversed due to the fluxoid \cite{haymaker}. \\
In Fig.\ref{bild2} we show the electric field strength (left column of Fig.\ref{bild2}) and the curl of the monopole current (right column of Fig.\ref{bild2}) on the $Q\bar{Q}$-axis for charge distances $d=1a$, $d=2a$ and $d=3a$. The full line represents a symmetrized ``measurement'', where the electric field strength (the curl of the monopole current) is connected to the close charge. The dotted line depicts the result for Schwinger lines connecting the field strength (the curl of the monopole current) with the far charge. As expected the sign of the electric field strength is changing at the position of charges. In the case $d=1a$ the flux is confined almost entirely within the link connecting the two charges. Only $2\%$ of the total flux reach the anticharge on the long way through the lattice boundary. For a charge distance of $d=3a$ the electric flux as well as the curl of the monopole current decreases in the middle between the charges what can be interpreted as fluctuations of the fluxoid. We want to emphasize that the curl of the monopole current has the same sign along the whole axis; according to the definition (\ref{dMax}) it forms a solenoid of left handed currents after going back to Minkowski space. \\
As we discussed in connection with the Gauss law a triplet or octet Polyakov line is the source of a coulor electric flux in a certain direction in the eight-dimensional $su(3)$-algebra space. The electric flux is continuously ``rotated'' on its path to the anticharge due to the non Abelian nature of the group $SU(3)$. A demonstration of this effect is the lowering of the dotted line compared to the full line in Fig.\ref{bild2} showing the profile of the electric field and the curl of the monopole current along the $Q\bar{Q}$-axis. Another evidence for this effect can be obtained from the evaluation of (\ref{rel1}) for $x=x_-$ resulting in
\begin{eqnarray}
\label{rela}
\lefteqn{ \langle \, \mbox{Tr} \left\{ \left[ ga^2 \, \mbox{Div} E(x_-) \right]^{x_+} \, L(x_+) \right\} \mbox{Tr} L^\ast(\vec{x}_-)\, \rangle \, = -i \frac{1}{2} \, g^2 \,\langle \mbox{Tr}\left[ L^{x_-}(x_+)_{(tl)} L^\dagger(x_-) \right] \, \rangle = } \nonumber \\ 
&& = -i \frac{1}{2} \, g^2 \;\,\langle \; \mbox{Tr} \left[L^{x_-}(x_+) L^\dagger(x_-) \right] - \frac{1}{3} \mbox{Tr} L(\vec{x}_+)  \mbox{Tr} L^\ast(\vec{x}_-) \; \rangle \nonumber \\
&& = -i \frac{1}{2} \, g^2 \;\,\langle \; 3 \left[L^{x_-}(x_+) \otimes L^\ast(x_-) \right]_{singlet} - \frac{1}{3} \mbox{Tr} L(\vec{x}_+)  \mbox{Tr} L^\ast(\vec{x}_-) \; \rangle \nonumber \\
&& = -i \frac{4}{3} \, g^2 \;\,\langle \; \left[L^{x_-}(x_+) \otimes L^\ast(x_-) \right]_{singlet} - \frac{1}{8} \mbox{Tr} \left[L^{x_-}(x_+) \otimes L^\ast(x_-) \right]_{octet} \; \rangle.
\end{eqnarray}
The last expression turns out to be proportional to the difference of Wilson loops in the singlet and octet representation of a quark-antiquark system at distance $d = \mid \vec{x}_+ - \vec{x}_- \mid$. In Fig.\ref{lcor} we show the numerical results for the absolute value of the left hand side of (\ref{rela}) divided by
\begin{equation}
\label{rela1}
\langle \, \mbox{Tr} L(\vec{x}_+)  \mbox{Tr} L^\ast(\vec{x}_-)\, \rangle \, = \langle \, \left[L^{x_-}(x_+) \otimes L^\ast(x_-) \right]_{singlet} + \frac{8}{8} \mbox{Tr} \left[L^{x_-}(x_+) \otimes L^\ast(x_-) \right]_{octet} \; \rangle. 
\end{equation}
The divergence produced by the anticharge, the sink of the electric field, is an $su(3)$-vector which with increasing distance $d = \mid \vec{x}_+ - \vec{x}_- \mid$ loses its correlation with the direction defined by the Polyakov line matrix at $x_+$, the source of the electric field.\\
The connection between $\mbox{Div} E(x_-)^{x_+}$ and the $Q\bar{Q}$ potential in (\ref{rela}) together with the correlation of two Polyakov lines in (\ref{rela1}) gives the nice opportunity to extract the potential of a $Q\bar{Q}$-pair in the singlet and octet channel
\begin{figure}[ht]
\vspace{1cm}
\centerline{\input{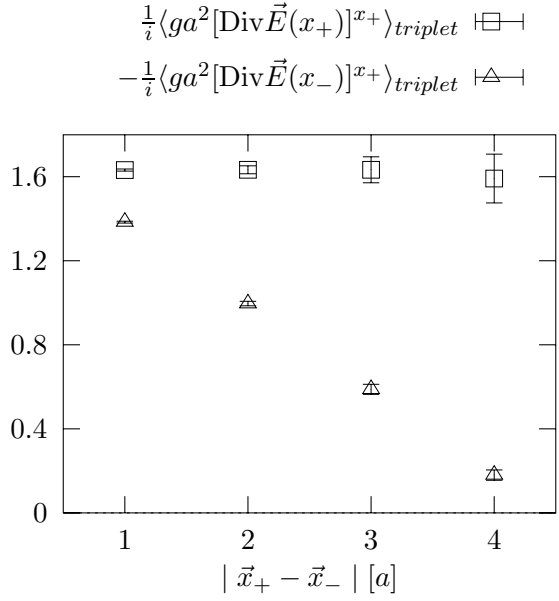}}
\vspace{0.3cm}
\caption{\label{lcor}The divergence of the electric field at the positions of a quark and an antiquark in dependence on the distance between the charges. In both cases the gauge dependent divergence is connected to the Polyakov line of the positive charge by a Schwinger line in order to define a gauge invariant quantity.} 
\end{figure}
\begin{eqnarray}
\label{potential}
V_{singlet}(\mid \vec{x}_+ - \vec{x}_- \mid) &=& \, -\frac{1}{N_ta} \, ln \, \langle \, \left[L^{x_-}(x_+)\otimes  L^\ast(x_-) \right]_{singlet} \rangle , \nonumber\\
V_{octet}(\mid \vec{x}_+ - \vec{x}_- \mid) &=& \, -\frac{1}{N_ta} \, ln \, \langle \, \frac{1}{8} \mbox{Tr} \left[L^{x_-}(x_+) \otimes L^\ast(x_-) \right]_{octet} \; \rangle,
\end{eqnarray}
with $N_t$ being the number of lattice sites in time direction. Because of the periodic boundary conditions on the $8^3 \times 2$-lattice the $Q\bar{Q}$-pair of (\ref{rela1}) may be connected by two different strings of length $r_1 = \mid \vec{x}_+ - \vec{x}_- \mid $ and $r_2 =N_xa-r_1$, where $N_x$ denotes the number of lattice sites in space direction. For charge distances smaller than $N_xa/2$ we can approximately neglect the contribution of the longer distance. For the distance $N_xa/2$ we have to consider only half of the value of the left-hand side of (\ref{rela1}) for the determination of the potentials in (\ref{potential}). The numerical results of (\ref{potential}) are shown in in Fig.\ref{potV1}.\begin{figure}[t]
\vspace{1cm}
\centerline{\input{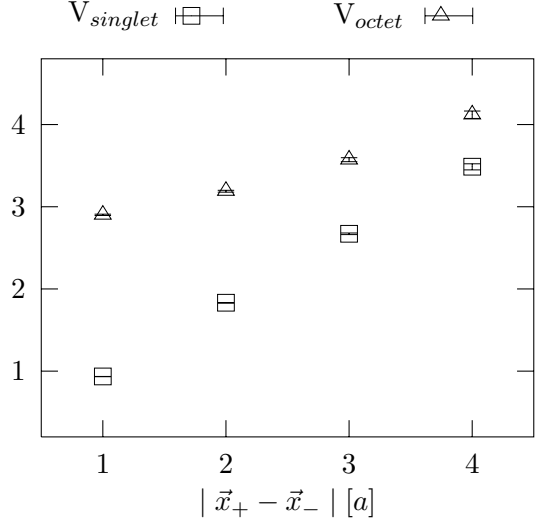}}
\vspace{0.3cm}
\caption{\label{potV1}$Q\bar{Q}$-pair potential in the singlet and octet channel as a function of the distance between the charges.} 
\end{figure}
\\
Since our aim is to test the dual superconductor picture of confinement, we turn now to the dual version\begin{figure}
\vspace{-1.5cm}
\centerline{\input{fig11.tex}}
\caption{\label{bild5}Ratio of the electric field strength and the curl of the monopole current on the links of the three-dimensional lattice for charge distance $d=1a$.}
\end{figure}
 of the London equation. Taking into account fluxoid contributions it reads \cite{haymaker} in Minkowski space\footnote[8]{According to our convention for the relation between Minkowski and Euclidean observables, in Euclidean space the dual London equation reads $\langle \vec{E} \rangle_{Q\bar{Q}} \; = \; -\lambda^2 \, \langle \mbox{curl} \vec{J}_m \rangle_{Q\bar{Q}}$.} 
\begin{equation}
\label{london}
\langle \vec{E} \rangle_{Q\bar{Q}} \; = \; \lambda^2 \, \langle \mbox{curl} \vec{J}_m \rangle_{Q\bar{Q}} \, + \, \mbox{fluxoid contributions}
\end{equation}
where $\lambda$ is the London penetration depth. In Fig.\ref{bild5} and in Fig.\ref{bild6} we show the ratio of the electric field strength and the curl of the monopole current on the links of the three-dimensional lattice for charge distances $d=1a$ and $d=2a$, where only results from symmetrized calculations are considered. It is clearly seen that the ratio
\begin{equation}
\label{lon2}
\tilde{\lambda}^2 \; := \; \langle E_i \rangle_{Q\bar{Q}}  / \langle (\mbox{curl} \vec{J}_m)_i \rangle_{Q\bar{Q}}
\end{equation}
rises with the distance $d$. (Be aware of the different scale in Fig.\ref{bild5} and in Fig.\ref{bild6}.) The same increase for $\tilde{\lambda}^2$ has been found in compact QED \cite{Zach} and can be interpreted as increasing effect of string fluctuations. For the case $d=1a$, off axis $\tilde{\lambda}^2$ lies in the range $\tilde{\lambda}^2=0.08\pm0.02$. This leads to a fluxoid on the link between the charges $\Phi=iga^2(\langle E_i \rangle_{Q\bar{Q}}-\tilde{\lambda}^2 \langle (\mbox{curl} \vec{J}_m)_i \rangle_{Q\bar{Q}})=1.604\pm0.077$ which agrees quite well with the theoretical value of $\Phi = \frac{4}{3} \, g^2 = 1.633$. For a charge distance $d=2a$ we obtain the value $\tilde{\lambda}^2=0.14\pm0.03$. The value of the corresponding fluxoid is then given by $\Phi=1.504\pm0.106$. A fit in an effective model allowing string fluctuations has shown that in compact QED \cite{Zach} the London penetration depth $\lambda$ is even smaller than $\tilde{\lambda}$ for distance $d=1a$. The above results show that the ratio $\langle E_i \rangle_{Q\bar{Q}}  / \langle (\mbox{curl} \vec{J}_m)_i \rangle_{Q\bar{Q}}$ behaves analogously to compact QED. According to our opinion this demonstrates the similarity of the confinement mechanism in QCD and compact QED.
\begin{figure}
\centerline{\input{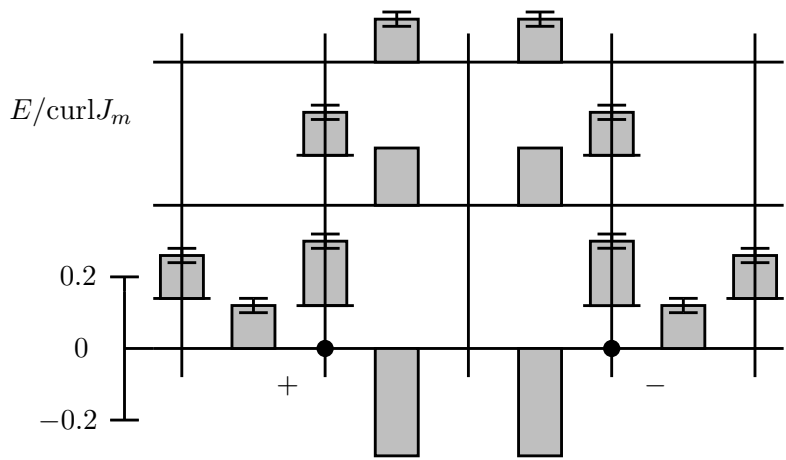}}
\caption{\label{bild6}Ratio of the electric field strength and the curl of the monopole current on the links of the three-dimensional lattice for charge distance $d=2a$.}
\end{figure}

\section{Conclusion}

In this article we developed a formalism how to determine the electric field distribution and the distribution of the curl of the monopole current of the gluonic vacuum with an external $Q\bar{Q}$-pair without using the technique of Abelian projection. We started with a derivation of the Gauss law for a colour triplet to find a well settled definition for the field strength on the lattice. By using the dual Maxwell equations we defined colourmagnetic monopoles and further the curl of the monopole current. We introduced Schwinger lines to determine the colourelectric field strength and the curl of the monopole current as gauge invariant quantities in lattice simulations and found that the ratio $\langle E_i \rangle_{Q\bar{Q}} / \langle (\mbox{curl} \vec{J}_m)_i \rangle_{Q\bar{Q}}$ behaves as predicted by the dual superconductor picture of confinement.\\
In future investigations we would like to measure the density of colourmagnetic monopoles defined in (\ref{dMax}). A candidate for a gauge invariant operator is the length of the monopole current in $su(3)$-space $\mid J_{\mu} \mid \, = \, \sqrt{\sum_{a=1}^8 J_{\mu}^aJ_{\mu}^a}$. First calculations show that the value of $\langle \mid J_{\mu} \mid \rangle $ decreases with increasing inverse coupling $\beta$ crossing the phase transition. But the quantity $\mid J_{\mu} \mid$ has the disadvantage that there are only positive contributions. Quantum fluctuations which are not of topological origin do not cancel. We want to point out that for the correlation function (\ref{Ecorrel}) the just mentioned problem does not occur: The Polyakov line fixes a direction in colour space, fluctuations of the operator of the electric field strength or the curl of the monopole current contribute equally with positive and negative sign and therefore cancel, whereas contributions of topological origin survive the averaging process.

\section*{Acknowledgements}

 We thank J.~Greensite for the hint to derive the Gauss law analytically, Yu.~A.~Simonov for critical comments concerning the validity of Bianchi identity and S.~Olejn{\'\i}k for interesting discussions.


\begin{thebibliography}{9}
%
\bibitem{Creutz} M. ~Creutz, {Phys.\ Rev.\/} {D21} (1980) 2308.
\bibitem{sommer} R.~Sommer, \np\ {B291} (1987) 673;\\
 W.~Feilmair, H.~Mar\-kum, \np\ {B370} (1992) 299;\\
 Y.~Peng, R.~W.~Haymaker, \pr\ {D47} (1993) 5104.
\bibitem{Hooft} G.~'t Hooft, in {\it High Energy Physics}, EPS International Conference, Palermo 1975, ed. A. Zichichi.
\bibitem{mandelstam} S.~Mandelstam, {Phys.\ Rep.\/} {23C} (1976) 245.
\bibitem{degrand} T.~A.~DeGrand, D.~Toussaint, \pr\ {D22} (1980) 2478.
\bibitem{Zach} M.~Zach, M.~Faber, W.~Kainz, P.~Skala, Phys. Lett. B358 (1995) 325.
\bibitem{haymaker} V.~Singh, R.~W.~Haymaker, D.~A.~Browne, \pr\ {D47} (1993) 1715.
\bibitem{thooft} G.~'t Hooft, Nucl.~Phys.~B190 (1981) 455.
\bibitem{green1}  J.~Greensite, J.~Iwasaki, {Phys.\ Lett.\/} {B255} (1991) 415;\\
L.~Del Debbio, M.~Faber, J.~Greensite, \np\ {B414} (1994) 594.
\bibitem{Skala} Diploma thesis of P.~Skala, "Die duale Londongleichung in der gluonischen Gitter-QCD", April 1995, Institut f\"ur Kernphysik, Technische Universit\"at Wien, Austria; P.~Skala, M.~Faber, M.~Zach, Proc. of the Internat. Workshop on Non-Perturbative Approaches to QCD, July 1995, Trento, Italy, World Scientific, in press; Proc. of the 29th Internat. Symposium on the Theory of Elementary Particles, August 1995, Buckow, Germany, Nucl. Phys. B (Proc. Suppl.), in press.
\bibitem{stefan} A.~Di~Giacomo, M.~Maggiore, \v S.~Olejn{\'\i}k, \np\ {B347} (1990) 441;\\ 
  P.~Cea, L.~Cosmai, \np\ B (Proc.~Suppl.) 42 (1995) 225.
\bibitem{charge} M.~M\"uller, M.~Faber, W.~Feilmair, H.~Mar\-kum, \np\ {B335} (1990) 502.
%
\end{thebibliography}
\end{document}